\documentclass[a4paper,11pt]{article}
\usepackage{epsfig}

\newcommand{\art}[6]{{\sc #1,  \rm #2 \it #3 \bf #4 \rm (#5), \mbox{#6}.}}

\newcommand{\book}[3]{{\sc #1, \it #2, \rm #3.}}

\newtheorem{proposition}{Proposition}[section]
\newtheorem{theorem}[proposition]{Theorem}
\newtheorem{corollary}[proposition]{Corollary}
\newtheorem{lemma}[proposition]{Lemma}
\newtheorem{definition}[proposition]{Definition}
\newtheorem{example}[proposition]{Example}
\newenvironment{proof}{\noindent{\bf Proof:$\;$}}{\hfill${\bf QED}$\medskip}
\newtheorem{remark}[proposition]{Remark}
\newenvironment{ams}{ {\small\it AMS Subject Classification} : }{}

\let\Definition=\definition
\renewcommand{\definition}{\Definition \rm}
\let\Example=\example
\renewcommand{\example}{\Example \rm}
\let\Remark=\remark
\renewcommand{\remark}{\Remark \rm}

\begin{document}

\title{\huge Quasi-Lagrangian systems of Newton equations}
\author{Stefan Rauch-Wojciechowski \and Krzysztof Marciniak$^{1}$ \and Hans Lundmark}
\date{Department of Mathematics, Link\"{o}ping University\\
      S-581 83 Link\"{o}ping, Sweden\\
      strau@mai.liu.se, krmar@mai.liu.se, halun@mai.liu.se \\ 
             September 22, 1999}

\maketitle

%1998 PACS No. 03.20+i, 02.30Hq, 02.30Jr

\begin{ams}
 58F07, 58F05, 70H20, 70D99, 35Q58
\end{ams}

\footnotetext[1]{ On leave of absence from Department of Physics, 
  A. Mickiewicz University, Pozna\'{n}, Poland}

\thispagestyle{empty}

\newpage

\begin{center}
       \large \bf Abstract
\end{center}

Systems of Newton equations of the form $\ddot{q}=-\frac{1}{2}A^{-1}(q)\nabla k$
with an  integral of motion quadratic in velocities are studied. These
equations generalize the potential case (when A=I, the identity matrix)
and they admit a curious quasi-Lagrangian formulation which differs
from the standard Lagrange equations by the plus sign between terms.
A theory of such quasi-Lagrangian Newton (qLN) systems having 
two functionally independent integrals of motion
is developed with focus on two-dimensional systems.
Such systems admit a bi-Hamiltonian formulation and are proved to
be completely integrable by embedding into five-dimensional
integrable systems. They are characterized by a linear, second-order
PDE which we call the fundamental equation.
Fundamental equations  are classified through
linear pencils of matrices associated with qLN systems.
The theory is illustrated by two classes of systems: 
separable potential systems and driven systems.
New separation variables for driven systems are found. These variables
are based on sets of non-confocal conics. An effective criterion for
existence of a qLN formulation of a given system is formulated
and applied to dynamical systems of the H\'{e}non-Heiles type.

\setcounter{section}{1}
\setcounter{proposition}{0}

\section*{I. Introduction}

\setcounter{equation}{0}
\def\theequation{1.\arabic{equation}}

In this paper we introduce and study such systems of
Newton equations $\ddot{q}=M(q)$ that can be generated
as equations of the form
\begin{equation}\label{ql}
0 = \frac{d}{dx}\frac{\partial E}{\partial \dot{q}}+\frac{\partial E}{\partial q}
\equiv \delta^+ E
\end{equation}
by an energy-like function quadratic in $\dot{q}$
\begin{equation}\label{ener}
E(q,\dot{q}) = \sum_{i,j=1}^{n}A_{ij}(q)\dot{q}_i\dot{q}_j + k(q) \equiv \dot{q}^tA\dot{q} + k(q),
\end{equation}
where $A(q)$ is an $n \times n$ symmetric matrix with real entries $A_{ij}(q)$. 
Here and in what follows we use the standard mechanical notation 
$q = (q_1,\ldots,q_n)^t$, $\dot{q}=(\dot{q}_1,\ldots,\dot{q}_n)^t$, for position
and velocity vectors (the superscript $t$ denotes the transpose of a matrix), where
$\dot{q}_k = \frac{\partial}{\partial x}q_k$, $k=1,\ldots,n$
with $x \in {\bf R}$ being the independent (time) variable.
By Newton equations we mean second order ordinary differential equations (ODE's)
of the form: acceleration $\ddot{q}$ is equal to the velocity independent
force $M(q)$. The force $M$ may be potential or not.

The equations in (\ref{ql}) are called here {\em quasi-Lagrangian} (qL) equations
since they differ from the Lagrange equations for $E(q,\dot{q})$ by sign
between terms only. These equations are shortly denoted $0 = \delta^+E
= (\delta_1^+E,\ldots,\delta_n^+E)^t$ where 
\begin{displaymath}
\delta^+_kE=\frac{d}{dx}\frac{\partial E}{\partial \dot{q}_k}
            + \frac{\partial E}{\partial q_k}
\end{displaymath}

The qL equations are not invariant with respect to arbitrary point transformation
but it can be easily shown that they remain invariant with respect to the affine 
change of variables $q = SQ+h$ where $Q=(Q_1,\ldots,Q_n)^t$ are the new variables and 
$S \in GL(n), h \in {\bf R}^n$. 

In the present article we shall mainly discuss quasi-Lagrangian sets of Newton
equations (qLN) generated by a function $E$ of the form (\ref{ener})
in the two-dimensional space of variables $q=(q_1,q_2)=(r,w)$. 
This class of equations (which seems to be completely new) is a very interesting
class because of its rich differential-algebraic structure and also because it contains
(as special cases)  the well understood class of point-separable potential Newton
equations $\ddot{q}=-\partial V(q)/\partial q$ and the class of non-potential Newton
equations of the triangular form $\ddot{r}=M_1(r,w),~\ddot{w}=M_2(w)$ which we shall 
call {\em driven}
systems. The qLN systems are not necessarily Lagrangian and thus they do not have any 
straightforward Hamiltonian formulation. 

In this paper we develop a theory
of completely integrable sets of qLN equations characterized by the existence
of two functionally independent integrals of motion quadratic in velocities: $E$ as above
and $F=\dot{q}^tB(q)\dot{q}+l(q)$. The existence of a second integral
of motion has far-reaching consequences; it eventually leads to wide classes of completely
integrable qLN systems.

\begin{example}\label{HDex}
The function 
$E=r\dot{r}\dot{w}-w\dot{r}^2-\alpha w r^2+\frac{1}{2}dr^2+\frac{w^2}{2r^4}$
when inserted into (\ref{ql}) gives rise to
\begin{displaymath} \renewcommand{\arraystretch}{1.5}
0 = \left[ \begin{array}{c} 
      \frac{d}{dx} \frac{\partial E}{\partial \dot{r}}
            + \frac{\partial E}{\partial r} \\
       \frac{d}{dx} \frac{\partial E}{\partial \dot{w}}
            + \frac{\partial E}{\partial w}
            \end{array} \right] = 
     \left[ \begin{array}{c}
     		-2w(\ddot{r}-\alpha r +\frac{w}{r^5})+r(\ddot{w}-4\alpha w +d) \\
     		r(\ddot{r}-\alpha r + \frac{w}{r^5})
     	    \end{array} \right] =
\end{displaymath}  
\begin{equation}\label{przyklad}
 = \left[ \begin{array}{cc} -2w & r \\
 				 r & 0 \end{array} \right]
        \left[ \begin{array}{c} 
        	\ddot{r}-M_1(r,w) \\
        	\ddot{w}-M_2(w) \end{array} \right] 				    
\end{equation}   
which is equivalent to a set of two Newton equations
\begin{equation}\label{twonprzykl}
  \begin{array}{l}
    \ddot{r} = \alpha r -\frac{w}{r^5} \equiv M_1(r,w)\\
    \ddot{w} = 4 \alpha w -d \equiv M_2(w)
  \end{array} 
\end{equation}
since the matrix
\begin{displaymath}
      \left[ \begin{array}{cc} -2w & r \\
 			 r & 0 \end{array} \right]
\end{displaymath}
is nonsingular. We see that the operation 
$0 = \delta^+E$ generates {\em linear combinations}
of the Newton equations (\ref{twonprzykl}).
\end{example}
Equations (\ref{twonprzykl}) were discovered accidentally as a Newton parameterization
of the  second stationary flow of the Harry Dym hierarchy \cite{two_newton}:
\begin{eqnarray*}
0 = \left( \frac{1}{4}\partial^3-\alpha \partial \right) 
    \left( \alpha u^{-3/2}-\frac{5}{16}\dot{u}^2u^{-7/2}+\frac{1}{4}\ddot{u}u^{-5/2}  \right) \\
= \left(\frac{1}{4}\partial^3-\alpha \partial \right) \left(-r^5\ddot{r}-\alpha r^6 \right)
\end{eqnarray*}
(here $\partial = \partial/\partial x$) where we substituted 
 $u=r^{-4}$. The substitution $w = -r^5\ddot{r}+\alpha r^6$
gives the system (\ref{twonprzykl}). The particular
feature of (\ref{twonprzykl}) is that it is a driven system: the equation for $w$ can be solved
independently and then the solution $w(x)$ drives the equation for $r$. 

\setcounter{section}{2}
\setcounter{proposition}{0}

\section*{II. General properties of quasi-Lagrangian Newton systems}

\setcounter{equation}{0}
\def\theequation{2.\arabic{equation}}

Let us consider an $n$-dimensional qL system $0 = \delta^+E$ with (quadratic in velocities)
energy-like function
\begin{equation}\label{energyn}
E(q,\dot{q}) = \sum_{i,j=1}^{n}A_{ij}(q)\dot{q}_i\dot{q}_j + k(q)
\end{equation}
with a symmetric (which can be assumed without loss of generality) matrix $A(q)=A^t(q)$. 
We shall formulate the necessary and sufficient condition
for the matrix $A(q)$ to make the equations $0 = \delta^+E$ equivalent to the set of equations
\begin{equation}\label{newton}
0 = \ddot{q}-M(q)
\end{equation}
with a velocity independent force $M(q) = (M_1(q),\ldots,M_n(q))^t$.

\begin{theorem}\label{rownowaznosc}
For the function $E$ given by (\ref{energyn}) with a nonsingular matrix $A(q)$ 
the following conditions are equivalent:
 \begin{enumerate}
  \item The equations $0 = \delta^+E$ are equivalent to the set of Newton equations
        $\ddot{q}=M(q)$ with velocity independent forces 
        $M=-\frac{1}{2}A^{-1}(q) \nabla k(q)$.
   \item The function $E$ is an integral of motion for the qL system $0 = \delta^+E$
   \item The matrix elements $A_{ij}(q)$ satisfy the following set of ``cyclic'' differential
         equations
         \begin{equation}\label{cyclicn}
         0 = \partial_i A_{jk}(q)+\partial_j A_{ki}(q)+\partial_k A_{ij}(q)
          ~~{\rm for~all~} i,j,k=1,\ldots,n.
         \end{equation}
 \end{enumerate}
\end{theorem}
Throughout the whole article the symbol $\nabla$ denotes the gradient operator and
$\partial_i = \partial/\partial q_i$. Later on we will also use the notation
$\partial_{ij}=\partial^2/\partial q_i \partial q_j$.

Statement 2 of the above theorem explains the name ``energy-like'' for the function $E$.

\vspace{15pt}

\begin{proof}
Let us calculate the $i$-th equation in $0=\delta^+E$:
\begin{displaymath}
0 = \delta_i^+E = \frac{d}{dx}\frac{\partial E}{\partial \dot{q}_i}
    +\frac{\partial E}{\partial q_i} 
    = \frac{d}{dx} \left( 2 \sum_j A_{ij}(q)\dot{q}_j \right) 
      + \sum_{j,k} \partial_i A_{jk}(q)\dot{q}_j\dot{q}_k 
      +\partial_i k =
\end{displaymath} 
\begin{equation}\label{1-3}
= 2\sum_j A_{ij}(q)\ddot{q}_j +\partial_i k
  + \sum_{j,k} \left( 
  	\partial_i A_{jk}(q) + \partial_j A_{ki}(q) + \partial_k A_{ij}(q)
  		\right)\dot{q}_j\dot{q}_k.
\end{equation}
The last equality in (\ref{1-3}) is due to the symmetry of $A(q)$. Thus, clearly, 
$2A\ddot{q}+\nabla k = 0$ if and only if the equations (\ref{cyclicn}) are satisfied
and the equivalence of 1 and 3 is established. 

Let us now calculate the total derivative of $E$ with respect to $x$.
\begin{displaymath}
\dot{E}  = \sum_i \left( 2\sum_{j} A_{ij}\ddot{q}_j
	+\partial_i k \right) \dot{q}_i 
	+\sum_{i,j,k}\partial_k A_{ij}\dot{q}_i \dot{q}_j \dot{q}_k =
\end{displaymath}
\begin{equation}\label{Edot}
 = \sum_i \left( 2\sum_{j} A_{ij}\ddot{q}_j
    +\partial_i k \right) \dot{q}_i  +\frac{1}{3}\sum_{i,j,k}\left( \partial_i A_{jk}
    +\partial_j A_{ki}+\partial_k A_{ij} \right)\dot{q}_i\dot{q}_j\dot{q}_k.
\end{equation}
The second term on the right hand side of the above equation has been rewritten 
by renaming indices. It contains precisely the cyclic conditions (\ref{cyclicn}).
So, if one (and thus both) of the statements 1 and 3 are satisfied, then both terms in
(\ref{Edot}) vanish. On the other hand, if $\dot{E}=0$ then terms at different powers of
$\dot{q}_i$ in (\ref{Edot}) must be equal to zero, which implies
both 1 and 3.
\end{proof}

\begin{remark}\label{forn=2}
For n=2  the general solution of equations (\ref{cyclicn}) can easily be found. 
It is
\begin{equation}\label{coefficients2}
 \begin{array}{ccc}
   A_{11}(w)    &=& aw^2+bw+\alpha \\
   2A_{12}(r,w) &=& -2arw-br-cw+\beta\\
   A_{22}(r)    &=& ar^2+cr + \gamma
 \end{array}
\end{equation}
with some real constants $a,b,c,\alpha,\beta,\gamma$. The corresponding qLN equations
read explicitly as
\begin{displaymath} \renewcommand{\arraystretch}{1.5}
0 = \left[ \begin{array}{c} 
      \frac{d}{dx} \frac{\partial E}{\partial \dot{r}}
            + \frac{\partial E}{\partial r} \\
       \frac{d}{dx} \frac{\partial E}{\partial \dot{w}}
            + \frac{\partial E}{\partial w}
            \end{array} \right] = 
    	2 \left[ \begin{array}{cc} A_{11} & A_{12} \\
 				 A_{12} & A_{22} \end{array} \right]
 		\left[ \begin{array}{c} 
        	\ddot{r}-M_1(r,w) \\
        	\ddot{w}-M_2(r,w) \end{array} \right] 		
\end{displaymath}
where
\begin{eqnarray*}
  M_1(r,w)=\frac{1}{2\det(A)}\left(A_{12}\frac{\partial k}{\partial w} 
                     -A_{22}\frac{\partial k}{\partial r} \right) \\
  M_2(r,w)=\frac{1}{2\det(A)}\left(A_{12}\frac{\partial k}{\partial r} 
                     -A_{11}\frac{\partial k}{\partial w} \right).
\end{eqnarray*}
\end{remark}

The remaining part of this work is mostly devoted to the case when  qLN equations
$\ddot{q}=-\frac{1}{2}A^{-1}\nabla k$ generated by $E$ admit a {\em second}
(quadratic in velocities)
 integral of motion $F(q,\dot{q}) = \sum_{i,j=1}^n B_{ij}(q)\dot{q}_i\dot{q}_j
+ l(q) \equiv \dot{q}^t B(q) \dot{q} + l(q)$ which is linearly, and therefore functionally,
independent of $E$.

\begin{theorem}[qLN systems with two integrals] \label{mainn}
Let the qLN system of Newton equations 
\begin{equation}\label{original}
  0 = \delta^+E = 2A(\ddot{q}+\frac{1}{2}A^{-1}\nabla k),
\end{equation}
generated by the function $E(q,\dot{q})=\dot{q}^tA(q)\dot{q}+k(q)$, admit a second,
functionally independent quadratic integral of motion 
$F(q,\dot{q})=\dot{q}^tB(q)\dot{q}+l(q)$. Then
 \begin{enumerate}
   \item The matrix $B(q)$ has the same structure as the matrix $A(q)$ in the sense 
         that the coefficients $B_{ij}(q)$ of $B(q)$ satisfy the  set of cyclic 
         differential equations (\ref{cyclicn}). 
   \item If $\det(B) \neq 0$ then 
          \begin{equation}\label{neccond}
           A^{-1}\nabla k = B^{-1}\nabla l
          \end{equation}
         and so the qLN system 
         $0 = \delta^+F = 2B(\ddot{q}+\frac{1}{2}B^{-1}\nabla l)$ generates the same Newton
         equations as $E$.
   \item Any differentiable function $f(E,F)$ generates the same system of Newton equations
         (by $0=\delta^+f(E,F)$) as $E$ does. In particular, any linear combination 
         $\lambda E + \mu F$ generates the same system of Newton equations.
 \end{enumerate}
\end{theorem}
The statement 2 shows one of peculiar features of qLN systems: all
quadratic (in velocities) integrals of motion of a qLN system generate the same system
(see also sec. 151 in \cite{whittaker}).

\vspace{15pt}

\begin{proof}
The requirement $\dot{B}=0$ yields (cf. (\ref{Edot}))
\begin{displaymath}
 0 = \sum_i \left( 2\sum_{j} B_{ij}(q)\ddot{q}_j
	+\partial_i l \right) \dot{q}_i 
	+\sum_{i,j,k}\partial_k B_{ij}\dot{q}_i \dot{q}_j \dot{q}_k =
\end{displaymath}
\begin{equation}\label{Fdot}
= \sum_i \dot{q}_i \left(2B \left(-\frac{1}{2}A^{-1}\nabla k \right) + \nabla l \right)_i
 +\sum_{i,j,k}\partial_k B_{ij}\dot{q}_i \dot{q}_j \dot{q}_k 
\end{equation}
where the index $i$ at the vector expression containing matrices $B$ and $A^{-1}$ denotes its $i$-th
component. The equality is satisfied identically with respect to $\dot{q}$ and so both sums must be
separately equal to zero.
 It follows, that the $B_{ij}$ satisfy the cyclic conditions $\partial_iB_{jk} + cycl. =0$
and that $2B \left(-\frac{1}{2}A^{-1}\nabla k \right) + \nabla l = 0$. The latter yields precisely
the equation (\ref{neccond}) since we assumed $\det(B) \neq 0$. So the statements 1 and 2
are proved. 

The operator $\delta^+$ acts as differentiation
on the algebra of constants of motion, so that
\begin{displaymath}
    0 = \delta^+f(E,F) = 
    \frac{\partial f}{\partial E}\delta^+E+\frac{\partial f}{\partial F}\delta^+F =
\end{displaymath}
\begin{displaymath}
     = 2\left(\frac{\partial f}{\partial E}A+\frac{\partial f}{\partial F}B \right)
     \left(\ddot{q} -M \right)
\end{displaymath}
(where $M=-\frac{1}{2}A^{-1}\nabla k = -\frac{1}{2}B^{-1}\nabla l$) which proves the
statement 3 of the theorem.
\end{proof}

It is important to stress that the equation (\ref{neccond}) is the necessary and sufficient
condition for the equivalence of the qLN system (\ref{original})
and the qLN system generated by  $F=\dot{q}^tB(q)\dot{q}+l(q)$. 
This condition will be used later.

\setcounter{section}{3}
\setcounter{proposition}{0}

\section*{III. qLN equations in two dimensions}

\setcounter{equation}{0}
\def\theequation{3.\arabic{equation}}

We shall from now on restrict our considerations to the case $n=2$. We will use
the notation $q=(q_1,q_2)^t=(r,w)^t$. The case of arbitrary $n$ is studied in a 
separate paper \cite{hans}.

For $n=2$ Theorem \ref{mainn} contains two special cases which explain the connection
of our theory with classical results \cite{whittaker} about separable potential Newton
equations and with the class of driven systems where one of the Newton equations depends only on
a single variable $r$ or $w$ and can be solved on its own.

\begin{corollary}\label{qLNdlasep}
 Assume that the Newton equations
 \begin{equation}\label{newtcor}
   \ddot{r} = M_1(r,w)~,~\ddot{w}  = M_2(r,w)
 \end{equation}
generated by the integral $E=\dot{q}^tA(q)\dot{q}+k(q)$ (with the matrix $A$
given by (\ref{coefficients2})) as $0 = \delta^+E$ 
have a potential force: $M_1=-\partial V/\partial r, \, M_2=-\partial V/\partial w$.
Then the potential $V(r,w)$ satisfies the Bertrand-Darboux equation \cite{whittaker}
\begin{equation}\label{BD}
 \begin{array}{l}
0 = (V_{ww}-V_{rr})( -2arw-br-cw+\beta )+2V_{rw}(aw^2-ar^2+bw+\\
   -cr+\alpha - \gamma)+3V_r(2aw+b)-3V_w(2ar+c)
  \end{array}
\end{equation}
(where the indices at $V$ denote partial derivatives with respect to $r$ and $w$)
with the coefficients $a,b,c,\alpha,\beta,\gamma$
 being exactly the coefficients of the polynomials in entries
 of the matrix $A$ as given by (\ref{coefficients2}). This means that the Newton system
 (\ref{newtcor}) can be solved by separating variables in the related Hamilton-Jacobi
 equation (see \cite{whittaker}).
\end{corollary}

\begin{proof}
 If $M$ is potential, then according to Theorem \ref{mainn} 
 $M=-\frac{1}{2}A^{-1}\nabla k = -\nabla V$ and so $\nabla k = 2A \nabla V$. 
 The potential $V$ exists provided that 
 $\partial^2 k/\partial r \partial w = \partial^2 k/\partial w \partial r$.
 This yields exactly the Bertrand-Darboux equation (\ref{BD}) for $V$.
\end{proof}

\begin{remark}
The quantity $k(r,w)/\det(A)$ satisfies the same Bertrand-Darboux equation as 
the potential $V$. This result can be verified directly but it also follows from 
Theorem \ref{fundth} in the next section.
\end{remark}

\begin{remark}
 Let us emphasize that the Hamiltonian system
 \begin{displaymath}
  \dot{r}=s~,~\dot{w}=z~,~\dot{s}=-\frac{\partial V}{\partial r}
  ~,~\dot{z}=-\frac{\partial V}{\partial w}
 \end{displaymath}
 generated by a separable natural Hamiltonian $H=\frac{1}{2}(s^2+z^2)+V(r,w)$
 can be reconstructed as the qLN system
 $0 = \delta^+E = 2A(\ddot{q}+\frac{1}{2}A^{-1}\nabla k)$
 from its second integral of motion $E$. This is easy to see, since 
 the above Hamilton equations are equivalent to $\ddot{r}=-\frac{\partial V}{\partial r}$,
 $\ddot{w}=-\frac{\partial V}{\partial w}$.
\end{remark}

The second class of equations satisfying the assumptions of Theorem \ref{mainn}
is the class of qLN systems of the form
\begin{equation}\label{driven2}
 \ddot{r}=M_1(r,w)~,~\ddot{w}=M_2(w)
\end{equation}
which naturally generalizes the system in Example \ref{HDex}. Such systems are called
{\em driven} since the equation for $w$ can be solved independently and then $w(x)$ can be
substituted into the equation for $r$.
Observe that the second equation (and thus the whole system)
admits an extra integral of motion of the form $F(w,\dot{w})=\dot{w}^2/2-\int M_2(w)dw$.
The qLN system $0=\delta^+E$ attains the form (\ref{driven2}) if and only if the second
component $M_2$ of the force $-\frac{1}{2}A^{-1}\nabla k$ does not depend on r
\begin{equation}\label{drivencond}
\frac{\partial}{\partial r}(A^{-1} \nabla k)_2 =0.
\end{equation} 

\begin{example}
The qLN equations generated by the function
\begin{displaymath}
 E=r\dot{r}\dot{w}-w\dot{r}^2+k(r,w)
\end{displaymath}
are driven (i.e. have the form (\ref{driven2})) provided that $k(r,w)$ satisfies
the following second order PDE
\begin{displaymath}
 0 = \frac{\partial}{\partial r} \left( \frac{1}{r}k_r+\frac{2w}{r^2}k_w \right)
\end{displaymath}
which is a specialization of (\ref{drivencond}). 
The general solution of the above equation is
\begin{displaymath}
k(r,w)=f \left(\frac{r^2}{w}\right)+r^2g(w)
\end{displaymath}
with arbitrary twice differentiable functions $f$ and $g$. The corresponding
qLN system attains the form
\begin{displaymath}
  \ddot{r} = -rg'(w)+\frac{r}{w^2}f' \left( \frac{r^2}{w}\right)
   ~,~\ddot{w}=-2\frac{d}{dw} \left( wg(w) \right)
\end{displaymath}
and can be solved by quadratures (see Section 7).
The second integral of motion of our system,
$F=\dot{w}^2/2-\int M_2(w)dw = \dot{w}^2/2 + 2wg(w)$, yields the matrix 
$B$ \begin{displaymath}
 B = \left[ \begin{array}{cc} 0 & 0 \\ 0 & 1/2 \end{array} \right]
\end{displaymath}
which is singular so $F$ does not generate our system.
 However any linear combination $\lambda E + \mu F$
of $E$ and $F$ (with both $\lambda$ and $\mu$ $\neq 0$)
is another integral of motion with a non-singular matrix
$B'=\lambda A +\mu B$ generates the same driven system as $E$.
\end{example}

Existence of two functionally independent constants of motion does not 
automatically imply Liouville integrability since we also need a Hamiltonian
formulation for our equations of motion. Our systems usually do not have a
Lagrangian formulation and so they do not have the standard Hamiltonian formulation.
On the other hand the special system discussed in Example \ref{HDex}, being
a stationary flow of the Harry Dym hierarchy, is expected to be integrable.
The question thus arises if/when our qLN systems 
 possess a non-standard Hamiltonian formulation. In Section 6
we shall demonstrate the existence of new Poisson structures for qLN systems
and their close relationship with Poisson pencils for separable potentials.
We shall also explain there when and in what sense our qLN systems
are integrable.

\setcounter{section}{4}
\setcounter{proposition}{0}

\section*{IV. Fundamental equation}

\setcounter{equation}{0}
\def\theequation{4.\arabic{equation}}

We shall now characterize those two-dimensional qLN systems which admit two (quadratic
in velocities) functionally independent integrals of motion $E$ and $F$, with the force
$M=-\frac{1}{2}A^{-1}\nabla k = -\frac{1}{2}B^{-1}\nabla l$.
We remind the reader that for $n=2$ we use the notation $q=(q_1,q_2)^t=(r,w)^t$.

Let us consider two symmetric $2 \times 2$ matrices $A(r,w)$ and $B(r,w)$ both 
satisfying the cyclic conditions  (\ref{cyclicn}). According to  Remark \ref{forn=2} 
they must have the following structure
\begin{equation}\label{AiB}
A = \left[ \begin{array}{cc} A_{11} & A_{12} \\ A_{12} & A_{22} \end{array} \right]
~,~B = \left[ \begin{array}{cc}
        B_{11} & B_{12} \\ 
        B_{12} & B_{22} \end{array} \right]
\end{equation}
with the polynomial entries given by (cf. (\ref{coefficients2}))
\begin{equation}\label{strukturaA}
 \begin{array}{l}
   A_{11}(w) = a_1w^2+b_1w+\alpha_1 \\
   2A_{12}(r,w) = -2a_1rw-b_1r-c_1w+\beta_1\\
   A_{22}(r) = a_1r^2+c_1r + \gamma_1
 \end{array}
\end{equation}
and
\begin{equation}\label{strukturaB}
 \begin{array}{l}
   B_{11}(w) = a_2w^2+b_2w+\alpha_2 \\
   2B_{12}(r,w) = -2a_2rw-b_2r-c_2w+\beta_2\\
   B_{22}(r) = a_2r^2+c_2r + \gamma_2
 \end{array}
\end{equation}
with some arbitrary real constants $a_1,\ldots,\gamma_2$.

\begin{theorem}[fundamental equation] \label{fundth}
Let
\begin{equation}\label{systemik}
\left[ \begin{array}{c} \ddot{r} \\ \ddot{w} \end{array} \right]
 = -\frac{1}{2}A^{-1}\nabla k = -\frac{1}{2}B^{-1}\nabla l,
\end{equation}
with nonsingular $2 \times 2$ matrices $A,B$ given by (\ref{AiB}),(\ref{strukturaA}) 
and (\ref{strukturaB}), be a set of qLN equations. Then the functions
$K_1=k/\det(A)$ and $K_2=l/\det(B)$ both satisfy the same linear, second order,
partial differential equation
\begin{equation}\label{fundeq}
\begin{array}{rcl}
0&=& 2(A_{12}B_{22}-A_{22}B_{12})\,K_{rr} \\
&-&  2(A_{11}B_{22}-A_{22}B_{11})\,K_{rw} \\
&+&  2(A_{11}B_{12}-A_{12}B_{11})\,K_{ww}  \\
&+& 3 (A_{12} \partial_r B_{22} - B_{12} \partial_r A_{22} +
       A_{22} \partial_w B_{11} - B_{22} \partial_w A_{11} )\, K_r  \\
&-& 3 (A_{11} \partial_r B_{22} - B_{11} \partial_r A_{22} +
       A_{12} \partial_w B_{11} - B_{12} \partial_w A_{11} )\, K_w  \\
&+& 3  ( \partial_r A_{22} \, \partial_w B_{11} -
         \partial_r B_{22} \, \partial_w A_{11} )\, K
\end{array}
\end{equation}
which explicitly reads
\begin{equation}\label{fundeqex} \arraycolsep 0pt 
 \renewcommand{\arraystretch}{1.1}
 \begin{array}{lrl} 
  0 \,  = \, & 2  K_{rr} & \left[  \gamma_2\beta_1-\gamma_1\beta_2
      + (b_2\gamma_1 - \gamma_2 b_1+ \beta_1 c_2  - c_1\beta_2)r   
      + (\gamma_1 c_2 - \gamma_2 c_1) w  \right. \\
    & & + \left. (  b_2 c_1 - c_2 b_1+  a_2\beta_1 -  a_1\beta_2) r^2
      + 2(\gamma_1 a_2-\gamma_2 a_1) w r \right. \\ 
    & & \left. + ( a_1 b_2 -  a_2 b_1) r^3+ ( a_2 c_1- c_2 a_1) w r^2 ) \right]  \\
    & +4  K_{rw} & \left[ \alpha_2\gamma_1 -\alpha_1\gamma_2 
      + ( \alpha_2 c_1-\alpha_1 c_2 ) r  
      + ( b_2\gamma_1 - \gamma_2 b_1) w   \right.\\      
    & &  \left.   +  ( \alpha_2 a_1-\alpha_1 a_2) r^2            
      + (\gamma_1 a_2-\gamma_2 a_1) w^2
      + ( b_2 c_1 -  c_2 b_1) rw 
      \right. \\ 
    & & \left.  + (a_1 b_2 - a_2 b_1) wr^2
      + ( a_2 c_1- c_2 a_1) rw^2 \right]\\
 & +2K_{ww} & \left[\alpha_1\beta_2 -\alpha_2\beta_1
      + (\alpha_2 b_1 - \alpha_1 b_2) r
      + (\alpha_2 c_1 - \alpha_1 c_2 +  \right. \\
    & & \left.  b_1\beta_2 - b_2\beta_1)w +  (a_1\beta_2 -  a_2\beta_1 +  b_2 c_1 - c_2 b_1) w^2
         \right. \\ 
    & & \left.   + 2(\alpha_2 a_1-\alpha_1 a_2)wr  + ( a_2 c_1- c_2 a_1) w^3
          + ( a_1 b_2 -  a_2 b_1) r w^2 \right] \\
 & +3K_r &\left[ 2 b_2\gamma_1- 2\gamma_2 b_1 + \beta_1 c_2 - c_1\beta_2 
      + (3 b_2 c_1-3 c_2 b_1  + 2 a_2\beta_1 - 2 a_1\beta_2) r \right. \\ 
  & & \left.
      + 4(\gamma_1 a_2 - \gamma_2 a_1) w 
      + 4(a_1 b_2- a_2 b_1 ) r^2  + 4( a_2 c_1 - c_2 a_1)rw \right] \\
 & +3K_w & \left[ 2\alpha_2 c_1- 2\alpha_1 c_2+  b_1\beta_2  -  b_2\beta_1
      + (2 a_1\beta_2 - 2 a_2\beta_1 + 3 b_2 c_1- 3 c_2 b_1  ) w \right. \\
    &  &+ \left. 4( \alpha_2 a_1-\alpha_1 a_2 ) r
      + 4( a_2 c_1 -  c_2 a_1) w^2  + 4(a_1 b_2- a_2 b_1) rw \right] \\
  &+6K & \left[ b_2 c_1 - c_2 b_1 + 2(a_1 b_2- a_2 b_1) r
      + 2( a_2 c_1 -  c_2 a_1) w \right]
\end{array}
\end{equation}
with $K$ denoting either $K_1$ or $K_2$ and 
$K_r = \partial K/\partial r$, $K_{rr}=\partial^2 K/\partial r^2$ and so on.

Conversely, any solution $K_2(q)$ of the equation (\ref{fundeq}) generates
 two  different systems of qLN equations
$\ddot{q}=-\frac{1}{2}A^{-1}\nabla k_1 = -\frac{1}{2}B^{-1}\nabla l_1$ and
$\ddot{q}=-\frac{1}{2}A^{-1}\nabla k_2 = -\frac{1}{2}B^{-1}\nabla l_2$, where the functions
$k_1,k_2,l_1,l_2$ are determined by the equations
\begin{equation}\label{generacja}
\begin{array}{ccc}
 l_1=K_2 \det(B) & ~ & \nabla k_1 = AB^{-1}\nabla(K_2 \det(B)) \\
 k_2=K_2\det(A) & ~ & \nabla l_2 = BA^{-1}\nabla(K_2 \det(A)) \end{array}
\end{equation}
\end{theorem}

We will call the equation (\ref{fundeq}) the {\em fundamental equation} associated with
the matrices $A$ and $B$.

The fundamental equation plays a crucial role in our  theory of qLN systems. 
Observe that it is invariant with respect to
the transformation $A \mapsto \lambda A + \mu B$, $B \mapsto \lambda' A + \mu' B$,
($\lambda, \lambda', \mu, \mu' \in {\bf R}$)
since the coefficients at every monomial in this equation are skew symmetric in $A$ and $B$. This
is consistent with statement 3 of Theorem \ref{mainn}, which asserts that
if any  pair $E,F$ of functions generates a qLN system then the linear combinations
$\lambda E + \mu F$ and $\lambda' E + \mu' F$  also generate the same system.
This explains that the assumption of nonsingularity for both $A$ and $B$ is nonessential
since if $\det(A) \neq 0$ a singular matrix $B$ can always be substituted by an 
invertible matrix $B'=\lambda A + \mu B$.
We shall investigate further  properties of the fundamental equation in the next theorem.

Notice that in the second part of Theorem \ref{fundth} one has to
reconstruct $l_2$ and $k_1$ by integrating 
the expressions for $\nabla l_2$ and for $\nabla k_1$. This can always be done, as the above
theorem implicitly states. Also, notice that in the fundamental equation
(\ref{fundeqex}) all terms of degree 4 and higher cancel so that the polynomial degree
of coefficients in this equation is less than or equal to 3.

\vspace{15pt}

\begin{proof} (of Theorem \ref{fundth})
Our qLN system (\ref{systemik}) is generated  by either of the two functions
$E(q,\dot{q}) = \dot{q}^tA\dot{q}+k$ and $F(q,\dot{q}) = \dot{q}^tB\dot{q}+l$
and so the condition (\ref{neccond}), i.e. $A^{-1}\nabla k = B^{-1}\nabla l$, must be satisfied.
This implies that $\nabla l = BA^{-1} \nabla k$. This equation for the function $l$ has solutions if and only if
its compatibility condition $l_{rw}=l_{wr}$ is satisfied. This yields a PDE for the function
$k$ which, after the substitution $k=K_1 \det(A)$ and with use of the cyclic conditions (\ref{cyclicn}),
yields that $K_1$ satisfies equation 
 (\ref{fundeq}). By inserting into this equation the explicit form
of the polynomials $A_{11},\ldots,B_{22}$ we obtain (\ref{fundeqex}). 
On the other hand, the condition
(\ref{neccond}) implies also $\nabla k = AB^{-1} \nabla l$, and its compatibility
condition $k_{rw}=k_{wr}$ gives a PDE which in terms of $K_2=l/\det(B)$ must attain
the form (\ref{fundeq}) with interchanged entries of $A$ and $B$
(since the equation $\nabla k = AB^{-1} \nabla l$  becomes $\nabla l = BA^{-1} \nabla k$ 
 when one exchanges $A,k$ and $B,l$).
Due to the skew-symmetry of coefficients of
the equation for $K_1$ with respect to the entries of matrices $A,B$
(clearly seen from the form of (\ref{fundeq}))
the obtained equation for $K_2$ differs from the equation for $K_1$
by a minus sign on the right-hand side only. This proves that $K_1$ and $K_2$ both satisfy (\ref{fundeq})
(notice, however, that this does {\em not} imply $K_1=K_2$).

The existence of $k_1$ (i.e. the possibility of integrating
the equations (\ref{generacja}) in order to obtain $k_1$) follows from the fact, that
the condition $\partial ^2 k_1/\partial r \partial w=\partial^2 k_1/\partial w \partial l$
together with $\nabla k_1 = AB^{-1}\nabla(K_2\det(B))$ yields precisely the fundamental equation
for $K_2$ which is satisfied due to assumptions. One can similarly  prove the existence of $l_2$.
The second statement of the theorem can now be proved by checking that both pairs $k_1,l_1$
and $k_2,l_2$ given by (\ref{generacja})
satisfy the condition (\ref{neccond}) and thus give rise to two
systems of qLN equations. 
\end{proof}

\begin{remark}\label{B=I/2}
 For $B(q)=\frac{1}{2}I$ (a $2 \times 2$ identity matrix) the equation (\ref{fundeq}) becomes 
the Bertrand-Darboux equation (\ref{BD}) characterizing all separable potentials 
since in this case $\ddot{q}=-\frac{1}{2}B^{-1}\nabla l= -\nabla l$ is a potential equation.
\end{remark}

The next theorem shows that  there exists a recursive relation between two different 
qLN systems constructed from a given solution $K_2(q)$ of the 
fundamental equation (\ref{fundeq}). This makes it possible to construct a doubly infinite sequence
of qLN systems corresponding to a given fundamental equation.

\begin{theorem}\label{recursionthm} {\bf (recursion theorem)}
Let $k_1,l_1$ and $k_2,l_2$ be two pairs of functions determined by a given solution $K_2$
of the fundamental equation (\ref{fundeq}) as in (\ref{generacja}). Then these functions
are related by the following linear algebraic equations
\begin{equation}\label{rekursja}
k_2=l_{1}\det(AB^{-1})~,~l_2=l_{1}{\rm Tr}(AB^{-1})-k_{1}
\end{equation}
(where {\rm Tr} denotes trace of matrix). Moreover, 
in the infinite sequence 
\begin{equation}\label{ciagrek}
 \cdots~~\begin{array}{cccc} 
   k_0 & ~ & ~ & ~ \\
    ~  & ~ &  ~  & \nearrow \\
    \downarrow  & ~ & K_1 & ~ \\
    ~  & \nearrow & ~ & ~ \\
   l_0 & ~ & ~ & ~ \end{array}
 \begin{array}{cccc} 
  k_1 & ~ & ~ & ~ \\
    ~  & ~ &  ~  & \nearrow \\
    \downarrow  & ~ & K_2 & ~ \\
    ~  & \nearrow & ~ & ~ \\
   l_1 & ~ & ~ & ~ \end{array}
 \begin{array}{cccc} 
   k_2 & ~ & ~ & ~ \\
    ~  & ~ &  ~  & \nearrow \\
    \downarrow  & ~ & K_3 & ~ \\
    ~  & \nearrow & ~ & ~ \\
   l_2 & ~ & ~ & ~ \end{array} \cdots
\end{equation}
of triples $(K_m,k_m,l_m)$, $m \in {\bf Z}$ defined recursively by
\begin{equation}\label{rekursjam}
k_m=l_{m-1}\det(AB^{-1})~,~l_m=l_{m-1}{\rm Tr}(AB^{-1})-k_{m-1}
\end{equation}
and by 
\begin{displaymath}
K_m=k_m/\det(A)=l_{m-1}/\det(B)
\end{displaymath}
the functions $k_m$ and $l_m$ satisfy $A^{-1}\nabla k_m = B^{-1}\nabla l_m$ and thus they both
determine the same (for a given $m$) qLN system 
$\ddot{q}=-\frac{1}{2}A^{-1}\nabla k_m = -\frac{1}{2}B^{-1}\nabla l_m$. 
All functions $K_m$
satisfy the fundamental equation (\ref{fundeq}) and are related through
the following two-step recursion
\begin{equation}\label{rekursjaK}
K_{m+1}=K_{m}{\rm Tr}(AB^{-1})-K_{m-1}\det(AB^{-1}).
\end{equation}
\end{theorem}
The above recursion is reversible.
The solution $K_m$ placed between $l_{m-1}$ and $k_m$ determines both
$l_{m-1}$ and $k_m$. The recursion (\ref{rekursjaK}) is soluble. Namely if we denote
the eigenvalues of the matrix $AB^{-1}$ by $\lambda_1$ and $\lambda_2$ then it can be proved that
for the case $\lambda_1 \neq \lambda_2$ the solution of (\ref{rekursjaK}) is
\begin{displaymath}
K_m=\frac{1}{\lambda_1 - \lambda_2}(K_1-\lambda_2 K_0)\lambda_1^m+
    \frac{1}{\lambda_1 - \lambda_2}(K_0\lambda_1- K_1)\lambda_2^m
\end{displaymath}
while in the case $\lambda_1 = \lambda_2$ the solution of (\ref{rekursjaK}) becomes
\begin{displaymath}
K_m=K_0\lambda_1^m+\left(\frac{K_1}{\lambda_1}-K_0\right)m\lambda_1^m.
\end{displaymath}
In both cases $K_0$ and $K_1$ are two subsequent solutions of the fundamental equation in the
sequence (\ref{ciagrek}) which are related by
\begin{displaymath}
\nabla \left( K_1 \det(B) \right) = BA^{-1}\nabla \left(K_0\det(A) \right).
\end{displaymath}

In order to prove the recursion theorem we need the following lemma.

\begin{lemma}\label{Xprop}
Let $X=AB^{-1}$ with matrices $A,B$ as above. Then
\begin{displaymath}
 X^{-1}\nabla (\det(X))=\nabla({\rm Tr}(X))
\end{displaymath}
\end{lemma}
This lemma follows from the cyclic properties (\ref{cyclicn}) of matrices $A$ and $B$
by a lengthy but straightforward calculation.

\begin{proof} (of the recursion theorem)
Consider a solution $K_2$ of the fundamental equation and the functions $k_1,l_1;k_2,l_2$ defined 
by (\ref{generacja}). Then obviously $k_2/\det(A)=l_1/\det(B)$ which immediately implies
$k_2=l_1\det(AB^{-1})$. Let $X=AB^{-1}$. Then
\begin{displaymath}
\begin{array}{l}
\nabla l_2 - \nabla(l_1{\rm Tr}(X)-k_1) = \\
= X^{-1}\nabla(K_2\det(A))-\nabla(K_2\det(B){\rm Tr}(X))+X\nabla(K_2\det(B))=\\
= X^{-1}\nabla(K_2\det(A))-({\rm Tr}(X)I-X)\nabla(K_2\det(B))-K_2\det(B)\nabla({\rm Tr}(X)) = \\
= X^{-1}\nabla(K_2\det(A))-X^{-1}\det(X)\nabla(K_2\det(B))-K_2\det(B)\nabla({\rm Tr}(X)) = \\
= K_2\det(B)\left(X^{-1}\nabla(\det(X))-\nabla({\rm Tr}(X)) \right) = 0
\end{array}
\end{displaymath}
where we used that $X^2-{\rm Tr}(X)X+\det(X)I=0$ as follows from the Cayley-Hamilton theorem.
The last equality is due to Lemma \ref{Xprop} above. Thus $l_2=l_1{\rm Tr}(X)-k_1$ up to a
non-essential additive constant. This proves the first assertion of the theorem. 

If we now define the sequence $\{(k_m,l_m)\}$ via the recursive procedure (\ref{rekursjam})
then a simple induction argument shows that each pair $(k_m,l_m)$ satisfies the condition
(\ref{neccond}) and thus both $k_m$ and $l_m$ determine the same qLN system.
Moreover, each $K_m=k_m/\det(A)=l_{m-1}/\det(B)$ is a solution of the fundamental equation
as theorem (\ref{fundth}) states. Finally, to obtain (\ref{rekursjaK}) it is enough to insert the
formula  $K_m=k_m/\det(A)=l_{m-1}/\det(B)$ into the second equation in (\ref{rekursjam}).
\end{proof}

\begin{example}\label{jacobifam}
(cf. Remark \ref{B=I/2}) For $B=\frac{1}{2}I$ (the potential case)
the recursion (\ref{rekursja}) takes the form
\begin{displaymath}
k_2=4V_1\det(A)~,~V_2=2{\rm Tr}(A)V_1-k_1
\end{displaymath}
with $V_1=l_1$.
This is the separable case when (\ref{fundeq}) reduces to the Bertrand-Darboux equation. In the generic case,
i.e. when $a \neq 0$ in (\ref{coefficients2}) the matrix $A(q)$ can be reduced
(with the use of affine transformations $q=SQ+h$ with $S \in GL(2,{\bf R}), h \in {\bf R}^2$,
see also Section 5) to the form
\begin{displaymath}
A(q)=\left[ \begin{array}{cc} -q_2^2+\lambda_2 & q_1q_2 \\ 
            q_1q_2 & -q_1^2+\lambda_1 \end{array} \right]
\end{displaymath}
If we now start with the harmonic oscillator potential $V_1=\frac{1}{2}(q_1^2+q_2^2)$ then the condition
$\nabla V_1 = \frac{1}{2}A^{-1}\nabla k_1$ gives $k_1=\lambda_2q_1^2+\lambda_1q_2^2$ and the recursion
formulas specify to
\begin{displaymath}
 \renewcommand{\arraystretch}{1.3}
 \begin{array}{c}
  k_2=2(q_1^2+q_2^2)(\lambda_1\lambda_2-\lambda_2q_1^2-\lambda_1q_2^2) \\
 V_2 = \lambda_1q_1^2+\lambda_2q_2^2-(q_1^2+q_2^2)^2 
 \end{array}
\end{displaymath}
thus reproducing the potential of the Garnier system \cite{garnier}. It can be shown 
that the above formulas prolongate to the $n=2$ case of the recursion for the Jacobi family of
elliptic separable potentials \cite{jacobi}.
\end{example}

In order to explain the character of the recursion (\ref{ciagrek}) more completely
let us consider instead of the pair $(A,B)$ of cyclic matrices another pair $(A+\mu B,B)$ with
$\mu \in {\bf R}$. As it can be shown (see below) this pair determines the same fundamental equation
as the pair $(A,B)$ does.
By choosing a solution $K_2$ of the fundamental equation and the pair
$(A+\mu B,B)$ we arrive at a different qLN system
$\ddot{q}=M_\mu(q) =-\frac{1}{2}(A+\mu B)^{-1}\nabla (K_2\det(A+\mu B))$. It turns out
that the force $M_\mu(q)$ is a linear  combination of two neighboring forces in the sequence
(\ref{ciagrek}) generated by $K_2$.

\begin{lemma}
Let $A$ and $B$ be two $2 \times 2$ matrices satisfying the cyclic conditions (\ref{cyclicn})
and let $K$ be a solution of the fundamental equation associated with $A$ and $B$. Let also
$\mu \in {\bf R}$. Then
\begin{displaymath}
  \left(A+\mu B \right)^{-1}\nabla \left(K\det(A+\mu B) \right) =
  A^{-1}\nabla \left(K \det(A) \right) +\mu B^{-1} \nabla \left(K\det(B)\right)
\end{displaymath}
\end{lemma}
This lemma is a consequence of Lemma \ref{Xprop}. It says that a solution of a given
fundamental equation determines the force $M$ (and so the system of qLN equations)
up to {\em linear combinations} of two consecutive systems in the recursion
(\ref{ciagrek}).

As we have mentioned the matrices $A$ and $B$ uniquely determine
the fundamental equation. The choice of $A,B$ which generate a given fundamental
equation is however not unique since the pair $A'=\alpha A + \beta B, B'=\gamma A + \delta B$
determines the same equation. One can also ask, to what extent a given fundamental equation
determines the pair $(A,B)$. The precise relationship between pairs $(A,B)$ and the fundamental 
equation is explained in the following theorem.

\begin{theorem}\label{jednoznacznosc}
Let $(A,B)$ be a pair of linearly independent matrices $A,B$ satisfying the cyclic 
conditions (\ref{cyclicn}). Then there is a 1-1 relationship between the linear span
$\{ \lambda A + \mu B : \lambda,\mu \in {\bf R} \}$ of $A$ and $B$ and the fundamental
equation (\ref{fundeq}) i.e.
\begin{enumerate}
 \item Any two linearly independent matrices $A'=\alpha A + \beta B, B'=\gamma A + \delta B$
        determine the same fundamental equation as $(A,B)$ does.
\item If the pair $(A',B')$ determines the same fundamental equation as $(A,B)$ does,
       then the matrices $A'$ and $B'$ belong to the linear span $\{ \lambda A + \mu B\}$ of $A$ and $B$.
\end{enumerate}
\end{theorem}

\begin{proof}
An easy calculation shows that the fundamental equation associated with the matrices 
$A'=\alpha A + \beta B$ and $B'=\gamma A + \delta B$ differs from the fundamental equation
associated with the matrices $A$ and $B$ by the multiplicative factor $\alpha \delta - 
\beta \gamma$ on the right-hand side, i.e. by the non-zero determinant of the transformation between
 $(A,B)$ and $(A',B')$ and so it is in fact the same equation. This shows assertion
1 of the theorem. 

Assume now that the equation (\ref{fundeq}) is associated with a pair $(A,B)$. Consider the
vector $\vec{X} = (X_1,X_2,X_3)^t \in {\bf R}^3$ of the coefficients of (\ref{fundeq})
at the highest derivatives $K_{rr},K_{rw},K_{ww}$ respectively. Then
\begin{equation}\label{ilwekt}
  \begin{array}{c}
 X_1=A_{12}B_{22}-A_{22}B_{12}\\
 X_2=A_{22}B_{11}-A_{11}B_{22}\\
 X_3=A_{11}B_{12}-A_{12}B_{11}   \end{array}
\end{equation}
or 
\begin{equation}\label{ilwekt2}
\vec{X}=\vec{A} \times \vec{B}
\end{equation}
 where $\vec{A}=(A_{11},A_{12},A_{22})^t$ and $\vec{B}=(B_{11},B_{12},B_{22})^t$
are three-dimensional vectors depending on $r$ and $w$. Hence, for a fixed $(r,w)$ both vectors 
$\vec{A}$ and $\vec{B}$ are orthogonal to $\vec{X}$. The coefficients at $K_r$, $K_w$ and $K$ 
yield equations which are differential consequences of (\ref{ilwekt}) and so they 
do not impose any additional restrictions on $(A,B)$. Suppose now that there exist
matrices $A'$ and $B'$ satisfying the cyclic condition (\ref{cyclicn}) and
associated with the same fundamental equation. This means that
the equation (\ref{ilwekt2}) has
another solution i.e. that $\vec{X}=\vec{A'} \times \vec{B'}$ so that the vectors
$\vec{A'}$ and $\vec{B'}$ are orthogonal to $\vec{X}$ and in consequence
they are linear combinations of $\vec{A}$ and $\vec{B}$:
$\vec{A'}=\alpha \vec{A}+\beta \vec{B}, \vec{B'}=\gamma \vec{A}+\delta \vec{B}$ with some 
coefficients that may depend on $r$ and $w$. For the corresponding matrices it immediately 
follows that
\begin{displaymath}
 A' = \alpha A + \beta B,~~B'=\gamma A + \delta B.
\end{displaymath}
It remains to show that the coefficients $\alpha,\beta,\gamma,\delta$ in fact
do not depend on $r$ nor $w$. This can be shown by inserting the explicit
form (\ref{strukturaA}) and (\ref{strukturaB}) of entries of matrices $A$, $B$, 
$A'$ and $B'$ into (\ref{ilwekt}). This shows assertion 2 of the theorem.
\end{proof}

\setcounter{section}{5}
\setcounter{proposition}{0}

\section*{V. Affine inequivalent forms of fundamental equation}

\setcounter{equation}{0}
\def\theequation{5.\arabic{equation}}

In this section we are interested in characterizing all different types of 
two-di\-men\-sio\-nal qLN systems admitting two functionally independent integrals
of motion $E$ and $F$ which are quadratic in velocities, i.e. systems of the form
\begin{equation}\label{system5}
\ddot{q}=M=-\frac{1}{2}A^{-1}\nabla k = -\frac{1}{2}B^{-1}\nabla l
\end{equation}
where $M$ is the force of the system. Every such system
is described by a pair of matrices $A(q),B(q)$ satisfying the cyclic conditions
(\ref{cyclicn}) and by a pair of functions $k(q),~l(q)$ satisfying 
 $A^{-1}\nabla k = B^{-1}\nabla l$. We remind the reader that the functions $k/\det(A)$ and
$l/\det(B)$ satisfy the same fundamental equation with the coefficients 
completely determined by the matrix elements of $A$ and $B$.

Let us first consider how the qLN system
$0=\delta^+E=2A(\ddot{q}+\frac{1}{2}A^{-1}\nabla k)$
 transforms under the affine transformation of coordinates
\begin{equation}\label{aff}
 q=SQ+h, ~S \in GL(2,{\bf R}),~h \in {\bf R}^2
\end{equation}
where $Q=(Q_1,\ldots,Q_n)^t$. It is easy to see, that 
under the affine transformation (\ref{aff}) the generating function $E$ transforms
as
\begin{equation}\label{transE}
 E(q(Q),\dot{q}(\dot{Q}))=\dot{Q}^tS^tA(q(Q))SQ+k(q(Q)),
\end{equation}
where $q(Q)=SQ+h$ and so $\dot{q}(\dot{Q})=S\dot{Q}$. It can be shown by a direct verification, 
that the transformed matrix
\begin{equation}\label{transA}
A_Q(Q)= S^tA(q(Q))S
\end{equation}
in (\ref{transE}) also satisfies the cyclic conditions (\ref{cyclicn}) and therefore (\ref{transE})
generates a qLN system. This means that the qLN system
$0=2A(\ddot{q}+\frac{1}{2}A^{-1}\nabla k)$ is indeed {\em invariant} with
respect to the affine change of coordinates (\ref{aff}). 

Let us now consider the system (\ref{system5}). Using (\ref{transA}) one can prove that 
the fundamental equation associated with the pair $(A,B)$ of matrices is also invariant
with respect to the affine transfromations (\ref{aff}). This means, that we can simplify
this fundamental equation by performing an appropriate affine change of coordinates.
But Theorem \ref{jednoznacznosc} makes it possible to classify  fundamental
equations, and therefore the corresponding qLN systems, by classifying {\em pairs}
of matrices $(A,B)$. Instead of working with the coefficients of the fundamental
equation we can thus work with linear spans $\{ \lambda A + \mu B \}$ of $A$ and $B$.
Since the affine transformations do not change the polynomial degree of matrices $A,B$,
the set of all linear spans of $A$ and $B$ can be divided into affine inequivalent classes
corresponding to different polynomial degree of $A$ and $B$. Each equivalence class will be
represented by the algebraically simplest pair of matrices obtained by the use of affine
transformations and linear combining of matrices (since the latter leave the fundamental
equation unchanged, see above).

In order to be more precise we shall introduce some notation.
By $A^{(i)}$ $(i=0,1,2)$ we will denote all matrices $A$ which satisfy the cyclic 
  conditions (\ref{cyclicn}) and have the highest degree of polynomial entries equal to $i$.
So, for example, the general form of matrices in the class $A^{(1)}$ is
\begin{displaymath}
 \left[ \begin{array}{cc}
         bq_1+\alpha & -\frac{1}{2}bq_1-\frac{1}{2}cq_2+\frac{1}{2}\beta \\
         \ast & cq_1+\gamma \end{array} \right]
\end{displaymath}
with arbitrary constants (parameters) $b,c,\alpha,\beta,\gamma$. We will use
the symbol $\ast$ to denote matrix elements determined by the symmetry
of a given matrix. Moreover, by
 $[A^{(i)},B^{(j)}]$ $(i,j=0,1,2)$ we will denote the class of (non-ordered)
pairs $(A,B)$ of linearly independent 
matrices $A,B$ such that one of the matrices belongs to $A^{(i)}$ and the other to $B^{(j)}$. 
We have, of course, $[A^{(i)},B^{(j)}]=[A^{(j)},B^{(i)}]$ and so we have precisely six 
such classes. Obviously, all classes $[A^{(i)},B^{(j)}]$ are invariant with respect to 
the affine transformations (\ref{aff}).
Notice  that if $(A,B) \in [A^{(2)},B^{(2)}]$ we can kill the coefficient $a_2$ at the 
second degree monomials in $B$ by subtraction $B \mapsto B - \frac{a_2}{a_1}A$ so
every element of this class can be reduced  to an element in
$[A^{(2)},B^{(1)}]$. Thus we have to consider only five classes. It is easy to realize that
the five classes $[A^{(i)},B^{(j)}]$ with $i \geq j,~j < 2$ are invariant with respect
to the affine transformations (\ref{aff}) and with respect to taking linear combinations
of pairs $(A,B)$. These two operations can now be used to find
for every class a simple representing pair $(A,B)$ which has a minimal number of free 
parameters (a simple representative).
Consider for example the class $[A^{(2)},B^{(0)}]$. The general form of matrices belonging
to this class is
\begin{equation}\label{general}
  \begin{array}{c} 
 A=\left[  \begin{array}{cc} 
          a_1q_2^2+b_1q_2+\alpha_1 & -a_1q_1q_2-\frac{1}{2}b_1q_1-\frac{1}{2}c_1q_2+\beta_1/2 \\
          \ast & a_1q_1^2+c_1q_1+\gamma_1 \end{array}\right]  
       \\ ~~\\
 B = \left[ \begin{array}{cc} \alpha_2 & \beta_2/2 \\
                   \beta_2/2 & \gamma_2 \end{array} \right]      
   \end{array}
\end{equation}
Translation by the vector $h=-\frac{1}{2a_1}(c_1,b_1)^t$ kills $b_1$ and $c_1$ in the matrix
$A$. Since translations obviously preserve the form of $B$ the above pair of matrices attains the
form
\begin{displaymath}
 A=\left[  \begin{array}{cc} 
          a_1q_2^2+\alpha_1 & -a_1q_1q_2+\beta_1/2 \\
          \ast & a_1q_1^2+\gamma_1 \end{array}\right]~~,~~
 B = \left[ \begin{array}{cc} \alpha_2 & \beta_2/2 \\
                   \beta_2/2 & \gamma_2 \end{array}  \right]         
\end{displaymath}
with some new constants denoted by the same letters as in (\ref{general}). Further, the
transformation  $A \mapsto A - \frac{\beta_1}{\beta_2}B$ kills
the coefficient $\beta_1$ in $A$. In case when $\beta_2 =0$ we can still kill $\beta_1$
in $A$ by the linear transformation $q=SQ$ with
\begin{displaymath}
S=\left[ \begin{array}{cc} 
     \frac{-t\gamma_1-\beta_1/2}{\alpha_1+t\beta_1/2} & 1 \\
      1 & t \end{array} \right]
\end{displaymath}
where $t \in {\bf R}$ must be chosen so that $\alpha_1+t\beta_1/2 \neq 0$ and $\det(S) \neq 0$
which can be always done.
Finally, we can divide both matrices by $a_1$ and $2\alpha_2$
respectively. So a simple representative of the class
 $[A^{(2)},B^{(0)}]$ has the form
\begin{equation}\label{representant}
 A=\left[  \begin{array}{cc} 
          q_2^2+\alpha_1 & -q_1q_2 \\
          \ast & q_1^2+\gamma_1 \end{array}\right]~~,~~
 B = \left[ \begin{array}{cc} 1/2 & \beta_2/2 \\
                   \beta_2/2 & \gamma_2 \end{array}  \right]       
\end{equation}
with four essential parameters. Both separable and driven systems belong to this class
since $B=\frac{1}{2}I$ for separable systems and $B=\rm{diag}(1/2,0)$ for driven systems.

We perform a similar reduction for each class  $[A^{(i)},B^{(j)}]$, $i \geq j,~j < 2$. The 
results are presented below.

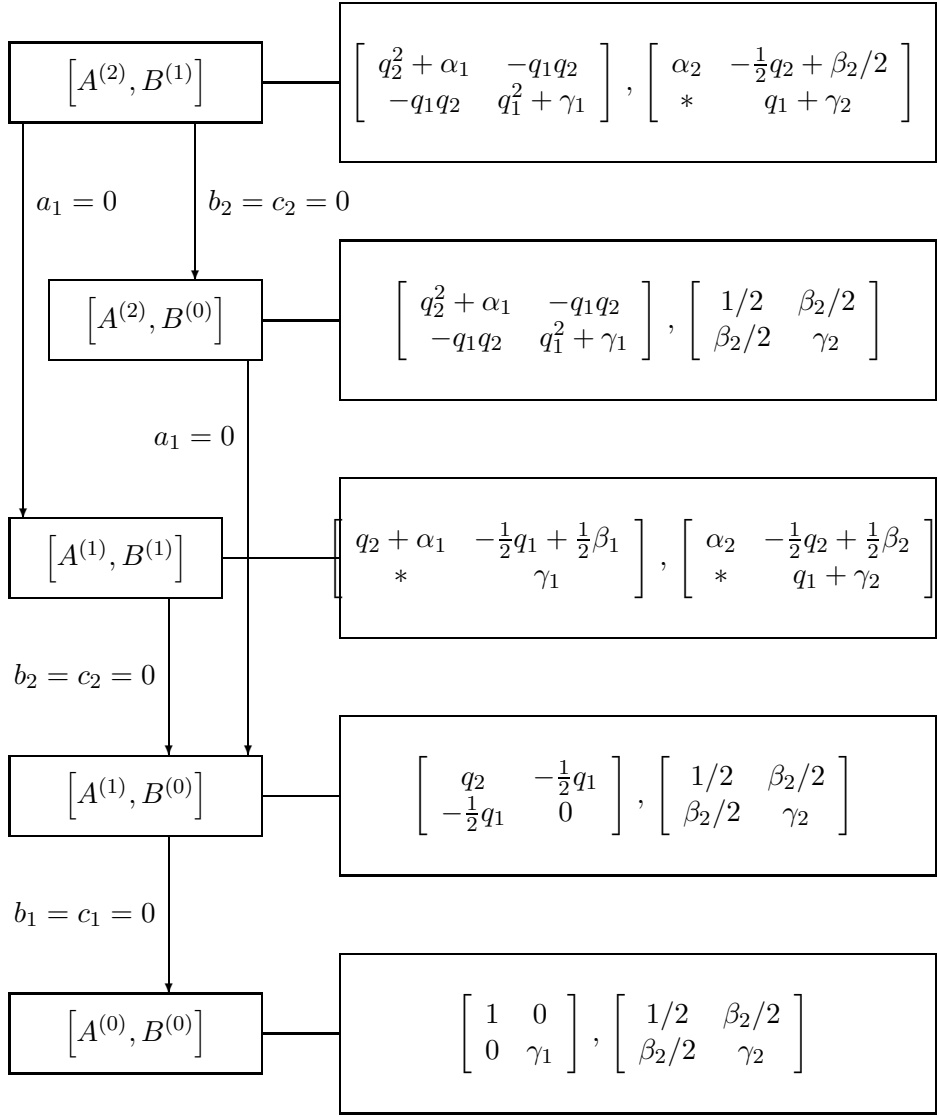
\begin{figure}[htpb]
\begin{picture}(345,440)(15,-20)
%KLASA A2B1
\put(0,365){\framebox(95,30){$\left[ A^{(2)},B^{(1)} \right]$}}
%teraz jej reprezentant
\put(125,350){\framebox(225,60){$ \left[ \begin{array}{cc}
                                q_2^2+\alpha_1 & -q_1q_2 \\
			        -q_1q_2 & q_1^2+\gamma_1
	 			\end{array} \right] \, , \,
				\left[ \begin{array}{cc}
				\alpha_2 & -\frac{1}{2}q_2+\beta_2/2 \\
                    			\ast  & q_1+\gamma_2
					  \end{array} \right]	$ }}
%teraz laczaca je linia
\put(95,380){\line(1,0){30}}
%KLASA A2B0 - wcinamy ja troche z lewej
\put(15,275){\framebox(80,30){$\left[ A^{(2)},B^{(0)} \right]$}}
%teraz jej reprezentant
\put(125,260){\framebox(225,60){$\left[ \begin{array}{cc}
				  q_2^2+\alpha_1 & -q_1q_2 \\
				  -q_1q_2 & q_1^2+\gamma_1
					\end{array} \right] \, , \,
				\left[ \begin{array}{cc}
				  1/2 & \beta_2/2 \\
				  \beta_2/2 & \gamma_2 \end{array} \right] $}}
%teraz laczaca je linia
\put(95,290){\line(1,0){30}}
%KLASA A1B1
\put(0,185){\framebox(80,30){$\left[ A^{(1)},B^{(1)} \right]$}}
%teraz jej reprezentant
\put(125,170){\framebox(225,60){$ \left[ \begin{array}{cc}
                                q_2+\alpha_1 & -\frac{1}{2}q_1+\frac{1}{2}\beta_1 \\
			        \ast & \gamma_1
					\end{array} \right] \, , \,
				\left[ \begin{array}{cc}
				\alpha_2 & -\frac{1}{2}q_2+\frac{1}{2}\beta_2 \\
                    			\ast  & q_1+\gamma_2
					  \end{array} \right]	$ }}
%teraz laczaca je linia
\put(80,200){\line(1,0){45}}
%KLASA A1B0
\put(0,95){\framebox(95,30){$\left[ A^{(1)},B^{(0)} \right]$}}
%teraz jej reprezentant
\put(125,80){\framebox(225,60){$ \left[ \begin{array}{cc}
                                q_2 & -\frac{1}{2}q_1 \\
			        -\frac{1}{2}q_1 & 0
					\end{array} \right] \, , \, 
				\left[ \begin{array}{cc}
				1/2 & \beta_2/2 \\
                    		\beta_2/2  & \gamma_2
					  \end{array} \right]	$ }}
%teraz laczaca je linia
\put(95,110){\line(1,0){30}}
%KLASA A0B0
\put(0,5){\framebox(95,30){$\left[ A^{(0)},B^{(0)} \right]$}}
%teraz jej reprezentant
\put(125,-10){\framebox(225,60){$ \left[ \begin{array}{cc}
                               1 & 0 \\
			        0 & \gamma_1
					\end{array} \right] \, , \,
				\left[ \begin{array}{cc}
				1/2 & \beta_2/2 \\
                    	        \beta_2/2  & \gamma_2
					  \end{array} \right]	$ }}
%teraz laczaca je linia
\put(95,20){\line(1,0){30}}
%SPECYFIKACJE
%z A2B1 do A1B1
\put(5,365){\vector(0,-1){150}}
\put(10,330){\makebox(0,0)[bl]{$a_1=0$}}
%z A2B1 do A2B0
\put(70,365){\vector(0,-1){60}}
\put(75,330){\makebox(0,0)[bl]{$b_2=c_2=0$}}
%z A2B0 do A1B0
\put(90,275){\vector(0,-1){150}}
\put(85,250){\makebox(0,0)[tr]{$a_1=0$}}
%z A1B1 do A1B0
\put(60,185){\vector(0,-1){60}}
\put(55,160){\makebox(0,0)[tr]{$b_2=c_2=0$}}
%z A1B0 do A0B0
\put(60,95){\vector(0,-1){60}}
\put(55,70){\makebox(0,0)[tr]{$b_1=c_1=0$}}
\end{picture}
\caption{Classification diagram}\label{diagramik}
\end{figure}

It is also easy to see that one can pass from one invariant class $[A^{(i)},B^{(j)}]$
to another by specifying values of free parameters. For example, by setting $a_1=0$ we obtain
$[A^{(1)},B^{(1)}]$ from  $[A^{(2)},B^{(1)}]$; by setting $b_2=c_2=0$ we get 
$[A^{(1)},B^{(0)}]$ and so on as shown in Figure \ref{diagramik}.
This figure presents --- for all classes  $[A^{(i)},B^{(j)}]$ --- complete results of simplification
of a generic pair $(A,B)$ belonging to each class
with the use of linear combinations and affine transformations.

%DIAGRAM

Below we list the form of the fundamental equation (\ref{fundeqex}) corresponding to 
the simple representative pair $(A,B)$ of each class as given in Figure \ref{diagramik}.
We use the notation $K_i=\partial K/\partial q_i$, $K_{ij}=\partial^2 K/\partial q_j \partial q_i$.

a) For $[A^{(2)},B^{(1)}]$
\begin{displaymath} \renewcommand{\arraystretch}{1.4}
 \begin{array}{cl}
   0 = & K_{11}\left(-\gamma_1\beta_2+\gamma_1q_2-2\gamma_2q_1q_2-\beta_2q_1^2-q_1^2q_2 \right)\\
   ~   & +2K_{12}\left( \alpha_2\gamma_1-\alpha_1\gamma_2-\alpha_1q_1
         -\gamma_2q_2^2+\alpha_2q_1^2-q_1q_2^2 \right) \\
   ~   & +K_{22}\left( \alpha_1\beta_2-\alpha_1q_2+2\alpha_2q_1q_2+\beta_2q_2^2-q_2^3 \right) \\
   ~   & +3K_1\left(-\beta_2q_1-2\gamma_2q_2-2q_1q_2\right)
         +3K_2\left(-\alpha_1+2\alpha_2q_1+\beta_2q_2-2q_2^2\right)\\
   ~   & -6q_2K \end{array}
\end{displaymath}

b) For $[A^{(2)},B^{(0)}]$
\begin{equation}\label{fea2c0} \renewcommand{\arraystretch}{1.4}
  \begin{array}{ccl}
 0 & = & K_{11}\left(-\gamma_1\beta_2-2\gamma_2q_1q_2-\beta_2q_1^2 \right) 
    +2K_{12}\left(\gamma_1-\alpha_1\gamma_2-\gamma_2q_2^2+q_1^2\right) \\
 ~ & ~ & +K_{22} \left( \alpha_1\beta_2+2q_1q_2+\beta_2q_2^2 \right)
   +3K_1\left(-\beta_2q_1-2\gamma_2q_2\right) \\
 ~ & ~ & +3K_2\left(2q_1+\beta_2w\right)
 \end{array}
\end{equation}

c) For $[A^{(1)},B^{(1)}]$
\begin{displaymath} \renewcommand{\arraystretch}{1.4}
 \begin{array}{cl}
   0 = & 2K_{11}\left(\gamma_2\beta_1-\gamma_1\beta_2+(-\gamma_2+\beta_1)q_1+q_2-q_1^2 \right)\\
   ~   & +4K_{12}\left( \alpha_2\gamma_1-\alpha_1\gamma_2-\alpha_1q_1
         -\gamma_2q_2-q_1q_2 \right) \\
   ~   & +2K_{22}\left( \alpha_1\beta_2-\alpha_2\beta_1+\alpha_2q_1+(\beta_2-\alpha_1)q_2-q_2^2 \right) \\
   ~   & +3K_1\left(-2\gamma_2+\beta_1-3q_1\right)
         +3K_2\left(-2\alpha_1+\beta_2-3q_2  \right)-6K \end{array}
\end{displaymath}

d) For $[A^{(1)},B^{(0)}]$
\begin{equation}\label{fea1c0}
0=2\gamma_2q_1K_{11}+4\gamma_2q_2K_{12}-2(q_1/2+\beta_2q_2)K_{22}+6\gamma_2K_1-3\beta_2K_2
\end{equation}

e) For $[A^{(0)},B^{(0)}]$
\begin{equation}\label{fea0c0}
0 = \gamma_1\beta_2K_{11}+2(\alpha_1\gamma_2-\alpha_2\gamma_1)K_{12}-\alpha_1\beta_2K_{22}
\end{equation}

What we present here is an illustrative characterization of different types
of fundamental  equations in terms of matrix pairs $(A,B)$. This provides a good intuitive
description of the world of qLN equations and helps to specify where two particular classes
--- separable potentials and driven systems --- belong. An alternative way of classifying qLN equations
with two quadratic integrals of motion is to simplify the fundamental equation (\ref{fundeq})
with the use of affine transformations as has been done for the Bertrand-Darboux equation 
\cite{whittaker} (see Example \ref{jakuproscicBD} below). This may amount to a similar picture
as we have presented above but the principles of simplification of the third-order polynomials
at $K_{rr}, K_{rw}$ and $K_{ww}$ are more difficult to discern. This is yet to be done.

\begin{example}\label{jakuproscicBD}
The classification  of types of the Bertrand-Darboux
equation with respect to Euclidean transformations leads to three forms of this equation
which are separable in either elliptic, parabolic or Cartesian coordinates.
According to 
Corollary \ref{qLNdlasep}, if a potential two-dimensional Newton system
\begin{displaymath}
 \ddot{q}_1=-\frac{\partial V}{\partial q_1}~,~\ddot{q}_2=-\frac{\partial V}{\partial q_2}
\end{displaymath}
with $V=V(q_1,q_2)$ possesses a second integral of motion of the form 
$E=\dot{q}^tA\dot{q}+k(q)$ with $A \in A^{(2)}$, then
it has the qLN form $\ddot{q}=-\frac{1}{2}A^{-1}\nabla k = -\frac{1}{2}B^{-1}\nabla V$ 
with $B=\frac{1}{2}I$ where $I$ is $2 \times 2$ identity matrix. Moreover the potential $V$
must satisfy the Bertrand-Darboux equation (\ref{BD}). This system belongs to the class 
$[A^{(2)},B^{(0)}]$ with $\beta_2=0$ and with $\alpha_2=\gamma_2=1/2$. The corresponding 
fundamental equation is exactly the Bertrand-Darboux equation since in this case
$K=V/\det(B)=4V$. The simplification procedure described above does not alter the form
of the matrix $B=\frac{1}{2}I$ and so the corresponding fundamental equation (\ref{fea2c0})
attains the form
\begin{equation}\label{niemamsily}
0 =  (V_{22}-V_{11})q_1q_2+V_{12}(q_1^2-q_2^2+\gamma_1-\alpha_1)-6q_2V_1+6q_1V_2
\end{equation}
which has only one essential parameter $\gamma_1 - \alpha_1$.
This form of the Bertrand-Darboux equation separates in the elliptic coordinates
\cite{whittaker}
\begin{displaymath}
 q_1=\frac{\xi \eta}{m}~,~q_2=\frac{1}{m}\left( (\xi^2-m^2)(m^2-\eta^2)\right)^{1/2}
\end{displaymath}
(where $\frac{1}{2}m^2=\alpha_1-\gamma_1$) which are its characteristic coordinates. In these
coordinates the Bertrand-Darboux equation (\ref{niemamsily}) takes the form
\begin{displaymath}
 0 = V_{\xi \eta}+\frac{2\eta}{\eta^2-\xi^2}V_{\xi}-\frac{2\xi}{\eta^2-\xi^2}V_{\eta},
\end{displaymath}
and its general solution is
\begin{displaymath}
 V(\xi, \eta)=\frac{f(\xi)-g(\eta)}{\xi^2 - \eta^2}
\end{displaymath}
with arbitrary functions $f$ and $g$.
The specification $a_1=0$ reduces the class $[A^{(2)},B^{(0)}]$ to $[A^{(1)},B^{(0)}]$.
The corresponding fundamental equation after the simplification procedure attains the form
(\ref{fea1c0}). In the case when $B=\frac{1}{2}I$ the final form of $B$ (after simplification) 
will be exactly the same (i.e. with $\beta_2=0,\alpha_2=\gamma_2=1/2$) and so the fundamental
equation (\ref{fea1c0}) reads
\begin{displaymath}
0 = q_1(V_{11}-V_{22})+2q_2V_{12}+3V_1
\end{displaymath}
It does not contain any parameters now. This equation separates
in the parabolic coordinates \cite{whittaker}
$q_1=\xi \eta$, $q_2=\frac{1}{2}(\xi^2-\eta^2)$
in which it takes the form
\begin{displaymath}
 0 = V_{\xi \eta}+\frac{2\eta}{\eta^2+\xi^2}V_{\xi}+\frac{2\xi}{\eta^2+\xi^2}V_{\eta}
\end{displaymath}
and has the solution
\begin{displaymath}
 V(\xi, \eta)=\frac{f(\xi)+g(\eta)}{\xi^2 + \eta^2}
\end{displaymath}
with arbitrary $f$ and $g$.
Further specification $b_1=c_1=0$ leads to the class $[A^{(0)},B^{(0)}]$ of constant symmetric
matrices. The corresponding fundamental equation in the course of simplification attains the
form (\ref{fea0c0}) which in the case $B=\frac{1}{2}I$ (again, this form of B survives the
simplification procedure - in this case just the diagonalization of $A$ by a rotation)
yields $V_{\xi \eta}=0$.
This is the case of the Bertrand-Darboux equation separable in 
(rotated) Cartesian coordinates.

The well known Bertrand-Darboux theorem (known also as the Whittaker theorem) 
\cite{whittaker,ankiewicz} says that potential Newton equations
admitting a second integral of motion quadratic in momenta admit separation in one of the
four coordinate systems: elliptic, parabolic, polar or Cartesian. The remaining polar coordinates
do not belong to our scheme, since radially symmetric potentials  $V(q_1^2+q_2^2)$
in the function $E=\frac{1}{2}(\dot{q}_1^2+\dot{q}_2^2)+V(q_1^2+q_2^2)$ have the angular momentum
$J=q_1\dot{q}_2-q_2\dot{q}_1$ as an integral of motion. This means that $F=J^2$ is the second
(quadratic in velocities) integral of motion of every potential system with radially symmetric potential.
But the function $F=J^2$ has $l(q) \equiv 0$ and therefore this case does not belong to our theory.
\end{example}

The above example indicates that the fundamental equation plays the same role in the theory of
 qLN equations as the Bertrand-Darboux equation does in the theory of separable potential forces
$M=-\partial V/\partial q$. For separable potentials the characteristic coordinates of the
Bertrand-Darboux equation determine the coordinates of separation which makes it possible
to solve the corresponding Newton equations by quadratures. 
In Section 7 we prove a similar result for the class of two-dimensional driven 
systems by showing that the characteristic coordinates of the fundamental equation
associated with a given driven system separate this system i.e. that in these coordinates
it is possible to integrate the system by quadratures. The question whether 
the characteristic coordinates of the fundamental equation
separate general qLN systems admitting two integrals of motion remains to be investigated.
We have here to do with a much broader theory depending on five essential parameters while the 
Bertrand-Darboux equation depends on one parameter only.

\setcounter{section}{6}
\setcounter{proposition}{0}

\section*{VI. Hamiltonian structures and complete integrability}

\setcounter{equation}{0}
\def\theequation{6.\arabic{equation}}

In this section we will establish a Hamiltonian formulation of two-dimensional qLN systems and discuss
their complete integrability. Let us consider first the qLN system 
$0=\delta^+E=2A(\ddot{q}+\frac{1}{2}A^{-1}\nabla k(q))$ generated by the function
$E=\dot{q}^tA(q)\dot{q}+k(q),~q=(q_1,q_2)^t$ with the $2 \times 2$ matrix $A(q)$ satisfying the
cyclic conditions (\ref{cyclicn}). This system usually does not have any Lagrangian formulation
and thus it does not have the standard Hamiltonian formulation. However we can always 
embed this system in a Hamiltonian qLN system in the five-dimensional phase space of variables
$(q_1,q_2,p_1,p_2,d)$ as the following theorem states.
\begin{theorem}[Hamiltonian form of qLN systems] \label{hamform}
Let
\begin{equation}\label{zanurz}
 0 = \ddot{q}+\frac{1}{2}A^{-1}(q)\nabla \left( k(q)+d\lambda \det(A(q)) \right)
\end{equation}
with $q=(q_1,q_2)^t$ be the qLN system generated by
\begin{displaymath}
  \hat{E}=\dot{q}^tA(q)\dot{q}+k(q)+d\lambda \det(A(q)) \equiv E + d\lambda \det(A)
\end{displaymath}
with some constant $\lambda$ and with $d \in {\bf R}$. Let also ${\cal M}$ be
 the extended 5-dimensional phase space of variables 
$(q_1,q_2,p_1,p_2,d)$ with $p_i=\dot{q}_i,~i=1,2$. Then the system (\ref{zanurz}) is equivalent to
\begin{equation}\label{hamzanurz} \renewcommand{\arraystretch}{1.3}
 \left[ \begin{array}{c} \dot{q} \\ \dot{p} \\ \dot{d} \end{array} \right]
 = \left[ \begin{array}{cc|c}
          0 & -\frac{\lambda}{2}G(q) & p \\
         \frac{\lambda}{2}G^t(q) & -\frac{\lambda}{2}F(q,p) & \hat{M}(q,d) \\ \hline
         -p^t & -\hat{M}^t(q,d) & 0 \end{array} \right] \nabla_{\cal \!\!M}d 
   \equiv \Pi_{A}\nabla_{\cal \!\!M}d
\end{equation}
where $\nabla_{\cal \!\!M}=\left(\partial/\partial q_1,\partial/\partial q_2,
           \partial/\partial p_1,\partial/\partial p_2,\partial/\partial d \right)^t$
is the gradient operator in ${\cal M}$ and where the $2 \times 2$ matrices $G$ and $F$
and the vector $\hat{M}$ are given by
\begin{displaymath}
 G(q)=\det(A)A^{-1} = \left[ \begin{array}{cc} A_{22} & -A_{12} \\
			-A_{12} & A_{11} \end{array} \right]
\end{displaymath}
\begin{displaymath}
 F_{12}(q,p)=\frac{1}{2}\left(\frac{\partial A_{22}}{\partial q_1}p_2
  -\frac{\partial A_{11}}{\partial q_2}p_1 \right),~~F=-F^t
\end{displaymath}
\begin{displaymath}
 \hat{M}(q,d)=M(q)-\frac{1}{2}d\lambda A^{-1}\nabla (\det(A))
\end{displaymath}
with $M(q)=-\frac{1}{2}A^{-1}\nabla k$ being the force of the qLN system  $0=\delta^+E$.
Moreover, the antisymmetric matrix $\Pi_{A}$ is Poisson and so (\ref{hamzanurz}) is the Hamiltonian
formulation of (\ref{zanurz}).
\end{theorem}
Notice that the matrix $G$ obtained above is symmetric due to the symmetry of $A$.

\begin{proof}
Since $\nabla_{\cal \!\!M}d=(0,0,0,0,1)^t$ the equation (\ref{hamzanurz}) yields $\dot{q}=p$, 
$\dot{p}=\hat{M}=-\frac{1}{2}A^{-1}\nabla \left(k+d\lambda \det(A) \right)$, $\dot{d}=0$,
i.e. it reproduces (\ref{zanurz}). The matrix $\Pi_A$ is antisymmetric and it is straightforward
to verify that it satisfies the Jacobi identity in the phase space ${\cal M}$.
\end{proof}

We remind the reader that the operator 
$\Pi : T^{\ast}{\cal M} \rightarrow T{\cal M}$
mapping fiberwise the cotangent bundle $T^{\ast}  \! {\cal M}$ of ${\cal M}$  into the tangent budle
$T \! {\cal M}$  is Poisson if the bilinear mapping 
$\{\cdot,\cdot\}_\Pi:C^{\infty}({\cal M}) \times C^{\infty}({\cal M}) \rightarrow C^{\infty}({\cal M})$
 defined for any pair of functions $f,g:{\cal M}\rightarrow {\bf R}$ by
\begin{displaymath}
\{f,g\}_\Pi=~<\nabla_{\cal \!\!M} f,\Pi\nabla_{\cal \!\! M} g>
\end{displaymath}
(where $<\cdot,\cdot>$ is the dual map between cotangent and tangent spaces of ${\cal M}$)
is a Poisson bracket.

\begin{remark}
 In the hyperplane $d=0$ the solutions of (\ref{hamzanurz}) coincide with the solutions
of $\ddot{q}=-\frac{1}{2}A^{-1}\nabla k$. Thus our original qLN system 
 $\ddot{q}=-\frac{1}{2}A^{-1}\nabla k$ is in a natural way embedded in the Hamiltonian system
(\ref{hamzanurz}).
\end{remark}

\begin{proposition}\label{casimirehat}
 The function
 \begin{displaymath}
  \hat{E}=p^tA(q)p+k(q)+d\lambda \det(A(q))
 \end{displaymath} 
is a Casimir function for the Poisson operator $\Pi_A$ in (\ref{hamzanurz}),
that is $\Pi_A\nabla_{\cal \!\!M}\hat{E}=0$.
\end{proposition}
One can check this proposition by a direct verification. 

A statement converse to the second statement of Theorem \ref{hamform} also holds.

\begin{theorem}\label{odwrotne}
 Let the antisymmetric matrix
 \begin{equation}\label{PiA} \renewcommand{\arraystretch}{1.4}
  \Pi= \left[ \begin{array}{cc|c}
          0 & -\frac{\lambda}{2}G(q) & p \\
         \frac{\lambda}{2}G^t(q) & -\frac{\lambda}{2}F(q,p) & \hat{M}(q,d) \\ \hline
         -p^t & -\hat{M}^t(q,d) & 0 \end{array} \right]
 \end{equation}
 be a Poisson operator in the space of variables $(q,p,d)$. Then
 \begin{enumerate}
  \item $G(q)$ must have the form
    \begin{equation}\label{Gform}
      G(q)=\left[ \begin{array}{cc}
           aq_1^2+cq_1+\gamma & aq_1q_2+\frac{b}{2}q_1+\frac{c}{2}q_2-\frac{\beta}{2} \\
           \ast & aq_2^2+bq_2+\alpha \end{array} \right]
    \end{equation}
       (thus it is symmetric) with some constants $a,b,c,\alpha,\beta,\gamma$ and so 
    \begin{equation}\label{wprowadzA}
     G=\left[ \begin{array}{rr} A_{22} & -A_{12} \\
    		-A_{21} & A_{11} \end{array} \right].
    \end{equation} 
  for some symmetric matrix $A(q)$ satisfying the cyclic conditions (\ref{cyclicn}).
  In other words, $\Pi=\Pi_A$ with $\Pi_A$ defined in (\ref{hamzanurz}) and with
  \begin{displaymath}
   A=\left[ \begin{array}{rr} G_{22} & -G_{12} \\ -G_{21} & G_{11} \end{array} \right].
  \end{displaymath}
  \item $F(q,p)$ must have the form
    \begin{equation}\label{Fform}
     F_{12}(q,p)=  \frac{1}{2}\left(\frac{\partial A_{22}}{\partial q_1}p_2
  -\frac{\partial A_{11}}{\partial q_2}p_1 \right),~~F=-F^t.
    \end{equation}
  \item $\hat{M}(q,d)$ must have the form
     \begin{displaymath}
      \hat{M}(q,d)=M(q)+d\lambda N(q)
     \end{displaymath}
  where $-2AM(q)=\nabla k$ for some function $k(q)$, so if $\det(G) \neq 0$ then 
  $M(q)=-\frac{1}{2}A^{-1}\nabla k$,  and where $N(q)=-\frac{1}{2}A^{-1}\nabla (\det(A))$.
 \end{enumerate}
\end{theorem}
\begin{proof}
The conditions $\{ \{ q_i,q_j \}_{\Pi} , q_k \}_{\Pi}+cycl=0$ and 
$\{ \{ q_i,q_j \}_{\Pi} , p_k \}_{\Pi}+cycl=0$ 
(where ``cycl'' means the cyclic permutation of expressions)
hold identically due to the block structure of $\Pi$. The condition
$\{ \{ q_i,q_j \}_{\Pi} , d \}_{\Pi}+cycl=0$ yields the symmetry of $G$: $G=G^t$. Further
\begin{displaymath}
0= \{ \{ q_i,p_j \}_\Pi , d \}_\Pi+cycl= -\frac{\lambda}{2}\left(p_1\frac{\partial G_{ij}}{\partial q_1}
 +p_2\frac{\partial G_{ij}}{\partial q_2}\right)+\frac{\lambda}{2}F_{ij}
 -p_i\frac{\partial \hat{M}_j}{\partial d}
\end{displaymath}
Let us denote the right hand side of the above equality by $-\frac{\lambda}{2}R_{ij}$. Notice that
$\partial \hat{M}_j/\partial d$ can not depend on $d$ and so we have
$\partial \hat{M}_j/\partial d = \lambda N_j(q)$ for some vector $N(q)=(N_1(q),N_2(q))^t$
which yields $\hat{M}(q,d)=M(q)+d\lambda N(q)$ for some vector $M(q)$. By taking linear
combinations of the conditions $R_{ij}=0$ and using the symmetry of $G$ and the antisymmetry of $F$
we get the following sets of equations
\begin{equation}\label{p1}
 \frac{\partial G_{11}}{\partial q_2}= \frac{\partial G_{22}}{\partial q_1}=0,~
 \frac{\partial G_{11}}{\partial q_1}=2N_1,~
  \frac{\partial G_{22}}{\partial q_2}=2N_2
\end{equation}
\begin{equation}\label{p2}
  \frac{\partial G_{12}}{\partial q_1}=N_2,~ \frac{\partial G_{12}}{\partial q_2}=N_1
\end{equation}
\begin{equation}\label{p3}
 F_{12}=p_2N_1-p_1N_2
\end{equation}
The equations (\ref{p1}) show that $G_{11}$ and $N_1$ depend only on $q_1$ and that
$G_{22}$ and $N_2$ depend only on $q_2$. The equations (\ref{p2}) give
$ \partial N_1/\partial q_1 = \partial^2 G_{12}/\partial q_1 \partial q_2 
= \partial N_2/\partial q_2$
and so all terms in this  expression must be equal to a constant $a$. Integration yields
\begin{equation}\label{p4}
 N_1=aq_1+c/2,~N_2=aq_2+b/2
\end{equation}
where $b$ and $c$ are integration constants. Substituting (\ref{p4}) into
(\ref{p1}) and (\ref{p2}) and integrating we get (\ref{Gform}). If we now introduce the 
symmetric matrix $A$ by the equality (\ref{wprowadzA}) and use (\ref{p1})
then (\ref{p3}) will attain the form (\ref{Fform}). 

It is straightforward to check that with the above forms of $F$ and $G$ the conditions
 $\{ \{ p_i,p_j \}_{\Pi} , p_k \}_\Pi+cycl=0$ and
 $\{ \{ q_i,p_j \}_\Pi , p_k \}_\Pi+cycl=0$ are satisfied identically.

Further, the condition  $\{ \{ p_1,p_2 \}_\Pi,d \}_\Pi+cycl=0$ after some calculations 
attains the form
\begin{displaymath}
 0=\frac{\partial}{\partial q_1}\left( G_{11}M_2-G_{21}M_1\right)
-\frac{\partial}{\partial q_2}\left( G_{22}M_1-G_{12}M_2\right)
\end{displaymath}
which means that in the vector
\begin{displaymath}
\left[ \begin{array}{c} G_{22}M_1-G_{12}M_2 \\-G_{21}M_1+G_{11}M_2 \end{array} \right]
= \left[ \begin{array}{rr} G_{22} & -G_{12} \\ -G_{21} & G_{11} \end{array} \right] 
\left[ \begin{array}{c} M_1 \\ M_2 \end{array} \right] = AM
\end{displaymath}
the mixed derivatives of its components are equal and so this vector is equal to the gradient
of some function $-\frac{1}{2}k(q)$, that is $AM=-\frac{1}{2}\nabla k$ or 
$M=-\frac{1}{2}A^{-1}\nabla k$. 

Finally, by direct calculation we verify that $N=-\frac{1}{2}A^{-1}\nabla (\det(A))$ and so
statement 3 of the theorem is proved.
\end{proof}

\begin{remark}
 This theorem generalizes the result of \cite{Paper5}. In particular, if we assume $M=-\nabla V(q)$
then we recover the known second Poisson operator for separable potential systems \cite{seppoiss}.
\end{remark}

Notice that $\hat{M}$ is the force of the 2-dimensional qLN system (\ref{zanurz}). This means that every
Poisson operator of the form (\ref{PiA}) is the Poisson operator for some qLN system 
of the form (\ref{zanurz}).

We are now in position to investigate complete integrability of qLN systems admitting
two quadratic, functionally independent integrals of motion. Notice first, that
Theorem \ref{odwrotne} provides us with an
alternative way of characterizing qLN systems generated by a quadratic integral of motion $E$:
by starting with a Poisson operator of the form (\ref{PiA}) we arrive at qLN systems generated by 
the Hamiltonian $H(q,p,d)=d$ which admit a quadratic integral $E$.
In a similar way the following theorem characterizes all qLN
systems admitting two independent quadratic integrals $E,F$.

\begin{theorem}[Poisson pencil] \label{pencilthm}
Consider the antisymmetric operator
\begin{equation}\label{pencil} \renewcommand{\arraystretch}{1.4}
  \Pi_{\mu}= \left[ \begin{array}{cc|c}
          0 & -\frac{\lambda}{2}G_{\mu}(q) & p \\
         \frac{\lambda}{2}G^t_\mu(q) & -\frac{\lambda}{2}F_{\mu}(q,p) & M(q)+d\lambda N_\mu(q) \\ \hline
         -p^t & -M^t(q)-d\lambda N^t_\mu(q) & 0 \end{array} \right]
\end{equation}
where 
\begin{displaymath}
G_{\mu}=\left[ \begin{array}{rr} A_{22} & -A_{12} \\ -A_{21} & A_{11} \end{array} \right]
-\mu \left[ \begin{array}{rr} B_{22} & -B_{12} \\ -B_{21} & B_{11} \end{array} \right]
\equiv G_A-\mu G_B,
\end{displaymath}
with both matrices $A$ and $B$ satisfying the cyclic conditions (\ref{cyclicn}), 
%where the $(1,2)$ matrix element in the matrix $F_\mu(q,p)$ has the form
\begin{displaymath}
 [F_{\mu}]_{12}= \frac{1}{2}\left(\frac{\partial (A_{22}-\mu B_{22})}{\partial q_1}p_2
  -\frac{\partial (A_{11}-\mu B_{11})}{\partial q_2}p_1 \right) \equiv [F_{A}]_{12}-\mu [F_{B}]_{12}
\end{displaymath}
(with  $F=-F^t, F_A=-F_A^t, F_B=-F_B^t$) and 
\begin{displaymath}
 N_{\mu}=-\frac{1}{2}A^{-1}\nabla \left( \det(A) \right)
  +\frac{1}{2}\mu B^{-1}\nabla \left( \det(B) \right) \equiv N_A - \mu N_B.
\end{displaymath}
Then $\Pi_{\mu}$ is Poisson if and only if
\begin{equation}\label{Mformpen}
M(q)=-\frac{1}{2}A^{-1}\nabla k = -\frac{1}{2}B^{-1}\nabla l
\end{equation}
for some functions $k(q)$ and $l(q)$. Moreover, if we let
\begin{displaymath}
 \Pi_{\mu}=\Pi_1-\mu \Pi_2 \equiv
\end{displaymath}
\begin{displaymath} \renewcommand{\arraystretch}{1.4}
 \equiv   \left[ \begin{array}{cc|c}
          0 & -\frac{\lambda}{2}G_A & p \\
          \frac{\lambda}{2}G^t_A& -\frac{\lambda}{2}F_A & M+d\lambda N_A \\ \hline
         -p^t & -M^t-d\lambda N^t_A & 0 \end{array} \right] -\mu
       \left[ \begin{array}{cc|c}
          0 & -\frac{\lambda}{2}G_B & 0 \\
         \frac{\lambda}{2}G^t_B & -\frac{\lambda}{2}F_B & d\lambda N_B \\ \hline
         0 & -d\lambda N^t_B & 0 \end{array} \right]
\end{displaymath}
then both operators $\Pi_1$ and $\Pi_2$ are Poisson and so $\Pi_\mu =\Pi_1 -\mu \Pi_2$ is
a Poisson pencil.
\end{theorem}

\begin{proof}
According to the proof of Theorem \ref{odwrotne} the matrix $\Pi_\mu$ satisfies all the Jacobi
identities except possibly for $\{ \{ p_1,p_2\}_{\Pi_\mu},d\}_{\Pi_\mu}+cycl=0$, 
since $\Pi_\mu$ differs from
$\Pi_{A-\mu B}=\Pi_A-\mu \Pi_B$ by the form of $M(q)$ only. Like in the proof
of Theorem \ref{odwrotne} we find that $\{ \{ p_1,p_2\}_{\Pi_\mu},d\}_{\Pi_\mu}+cycl=0$
 yields that the mixed
derivatives of the components of the vector $-2(A-\mu B)M$ are equal and so 
$-2(A-\mu B)M=\nabla (k-\mu l)$ for some functions $k(q)$ and $l(q)$. By comparing  
coefficients at different powers of $\mu$ we get $-2AM=\nabla k$ and $-2BM=\nabla l$ and thus
$M=-\frac{1}{2}A^{-1}\nabla k = -\frac{1}{2}B^{-1}\nabla l$.

Further, $\Pi_1=\Pi_A$ in the notation of Theorem \ref{odwrotne} so it is Poisson. Easy
calculation shows that $\Pi_2$ is Poisson too.
\end{proof}

The above theorem states that if $M(q)$ is the force of a qLN system admitting
two functionally independent integrals of motion then the matrix $\Pi_\mu$ is a Poisson
pencil. We will establish its Casimir function, which will be a polynomial in
$\mu$. This will lead to a bi-Hamiltonian chain containing the qLN system (\ref{zanurz}).
We will prove that this chain is 
completely integrable. In this way we will show that our original qLN system
$\ddot{q}=-\frac{1}{2}A^{-1}\nabla k=-\frac{1}{2}B^{-1}\nabla l$
can be naturally embedded in a completely integrable bi-Hamiltonian system.

\begin{proposition}
Suppose that  $\Pi_\mu$ is Poisson, i.e. that (\ref{Mformpen}) is satisfied. Then the  function
\begin{equation}\label{hmiu}
 H_\mu = p^t(A-\mu B)p+k-\mu l + d\lambda \det(A-\mu B)
\end{equation}
is a Casimir function for $\Pi_\mu$, i.e. $\Pi_\mu \nabla H_\mu =0$.
\end{proposition}

\begin{proof}
 This proposition is a consequence of Proposition \ref{casimirehat}. If we modify the matrix
$\Pi_A$ by substituting the matrix  $A$ by $A-\mu B$ and substituting $k$ with 
$k-\mu l$ we obtain the matrix
\begin{equation}\label{s1}  \renewcommand{\arraystretch}{1.4}
       \left[ \begin{array}{cc|c}
          0 & -\frac{\lambda}{2}G_\mu & p \\
         \frac{\lambda}{2}G^t_\mu & -\frac{\lambda}{2}F_\mu & \tilde{M}_\mu \\ \hline
         -p^t & -\tilde{M}^t_\mu & 0 \end{array} \right]
\end{equation}
where
\begin{displaymath}
 \tilde{M}_\mu= -\frac{1}{2}(A-\mu B)^{-1}\nabla (k-\mu l)
 -\frac{1}{2}d\lambda (A-\mu B)^{-1}\nabla \left(\det(A-\mu B) \right).
\end{displaymath}
Due to Proposition \ref{casimirehat} the function (\ref{hmiu}) is the Casimir of
(\ref{s1}). But (\ref{s1}) is in fact equal to $\Pi_\mu$ since
it can be verified that 
$-\frac{1}{2}(A-\mu B)^{-1}\nabla (k-\mu l)=-\frac{1}{2}A^{-1}\nabla k = -\frac{1}{2}B^{-1}\nabla l$
and that  %due to Lemma \ref{Xprop} 
$(A-\mu B)^{-1}\nabla \left(\det(A-\mu B) \right) = A^{-1}\nabla \left( \det(A) \right)$  
$-$ $\mu B^{-1}\nabla \left( \det(B) \right)$.
\end{proof}

Let us collect terms in $H_\mu$ at different powers of $\mu$
\begin{displaymath}
H_\mu = p^tAp+k+d\lambda \det(A) + \mu \left( -p^tBp-l-d\lambda Y \right) 
+ \mu^2 \left( d\lambda \det(B) \right)
\end{displaymath}
\begin{displaymath}
\equiv \hat{E} + \mu \hat{F} + \mu^2 \hat{H} 
\end{displaymath}
with $Y=B_{11}A_{22}+B_{22}A_{11}-2B_{12}A_{12}$. Then the above proposition gives:
\begin{displaymath} \renewcommand{\arraystretch}{1.4}
\begin{array}{cl}
 0 = & \Pi_\mu \nabla H_\mu 
    = (\Pi_1-\mu \Pi_2)\nabla \left( \hat{E}+\mu \hat{F}+\mu^2 \hat{H} \right) =\\
 ~ & =\Pi_1 \nabla \hat{E} + \mu \left( \Pi_1\nabla \hat{F}-\Pi_2 \nabla \hat{E} \right)
     + \mu^2 \left(\Pi_1\nabla \hat{H}-\Pi_2\nabla \hat{F} \right) -\mu^3\Pi_2\nabla \hat{H}.
\end{array}
\end{displaymath}
By equating to zero the coefficients at different powers of $\mu$ we obtain the following
bi-Hamiltonian chain:
\begin{equation}\label{biham}
 \begin{array}{rcl}
  \Pi_1\nabla\hat{E} & = & 0 \\
  \Pi_1\nabla\hat{F} & = & \Pi_2 \nabla\hat{E} \\
  \Pi_1\nabla\hat{H} & = & \Pi_2 \nabla\hat{F} \\
 		0    & = & \Pi_2 \nabla\hat{H} \end{array}
\end{equation}

\begin{theorem}
 The bi-Hamiltonian chain (\ref{biham}) is completely integrable, i.e. both non-trivial 
bi-Hamiltonian vector fields 
\begin{displaymath}
V_1=  \Pi_1\nabla\hat{F} = \Pi_2 \nabla\hat{E},~~~~~
V_2=  \Pi_1\nabla\hat{H} = \Pi_2 \nabla\hat{F}
\end{displaymath}
in (\ref{biham}) are completely integrable.
\end{theorem}

\begin{proof}
(modification of the proof of Liouville-Arnold theorem \cite{arnold})
 Consider the 2-di\-men\-sio\-nal ma\-ni\-fold
${\cal N} = \{ x\in {\cal M}:\hat{E}(x)=E_0,\hat{F}(x)=F_0,\hat{H}(x)=H_0 \}$
in ${\cal M}$. Poisson bra\-ckets of all pairs of $\hat{E},\hat{F},\hat{H}$ induced
by both structures $\Pi_0$ and $\Pi_1$ are equal to zero, since the functions
$\hat{E},\hat{F},\hat{H}$ all belong to the same bi-Hamiltonian chain. For instance
$\{ \hat{F},\hat{H} \}_{\Pi_1} = ~<\nabla \hat{F},\Pi_1\nabla \hat{H}> \; = \; \;$
$<\nabla \hat{F},\Pi_2\nabla \hat{F}>~ =\{\hat{F},\hat{F}\}_{\Pi_2} =0$ with the second equality 
being a 
consequence of the bi-Hamiltonian structure of $V_2$. It follows that the Lie bracket
$[V_1,V_2]$ of both vector fields $V_1$ and $V_2$ is equal to zero, $[V_1,V_2]=
[\Pi_1\nabla\hat{F},\Pi_1\nabla\hat{H}]=0$, since the mappings $\Pi_i\nabla$ ($i=1,2$) 
are Lie algebra homomorphisms between the Lie algebra of vector fields on ${\cal M}$ and
the Lie algebra of all smooth functions on ${\cal M}$ with the Lie bracket defined
by $[f_1,f_2]=\{f_1,f_2\}_{\Pi_i}$. Moreover 
$<\nabla\hat{E},V_1> \; = \; \;$ $<\nabla\hat{E},\Pi_2\nabla\hat{E}>~ =\{\hat{E},\hat{E}\}_{\Pi_2}=0$
and similarly $<\nabla\hat{F},V_1>~ = 
~<\nabla\hat{H},V_2>~ =0$ which proves that $V_1$ is
tangent to ${\cal N}$. In the same way one can show that $V_2$ is also tangent to ${\cal N}$.
Direct verification shows that $V_1$ and $V_2$ are linearly independent. We thus have a 
2-dimensional sub-manifold ${\cal N}$ in ${\cal M}$ 
equipped with a pair of linearly independent, commuting vector fields $V_1$ and $V_2$.
We can now apply the construction of Liouville-Arnold \cite{arnold} and  conclude that both
$V_1$ and $V_2$ are completely integrable.
\end{proof}

\begin{corollary}
The qLN system $\ddot{q}=M=-\frac{1}{2}A^{-1}\nabla k = -\frac{1}{2}B^{-1}\nabla l$ with two
linearly independent matrices $A$ and $B$ satisfying the cyclic conditions (\ref{cyclicn})
is completely integrable in the sense that the trajectories of the system
\begin{equation}\label{pierwotny}
\left[ \begin{array}{c} \dot{q} \\ \dot{p} \end{array} \right] = 
\left[ \begin{array}{c} p \\ M \end{array} \right]
\end{equation}
coincide on the hyperplane $d=0$ with the trajectories of the completely integrable 
5-dimensional system
\begin{equation}\label{polykacz}
\left[ \begin{array}{c} \dot{q} \\ \dot{p} \\ \dot{d} \end{array} \right]
=V_2=\Pi_1\nabla\hat{H}=\Pi_2\nabla\hat{F}.
\end{equation}
\end{corollary}

\begin{proof}
Consider the vector field $V_2$ from (\ref{biham}). Obviously
\begin{displaymath}
V_2=\Pi_1\nabla\hat{H}=\lambda\det(B)\Pi_1\nabla d+\lambda d \Pi_1\nabla \left( \det(B) \right)
\end{displaymath}
and so in the hyperplane $d=0$ we have
\begin{equation}\label{nad=0}
V_2|_{d=0}=\lambda \det(B)\left[ \begin{array}{c} p \\ M \\ 0 \end{array} \right]
\end{equation}
which means that the hyperplane $d=0$ is invariant with respect to the action of the vector
field $V_2$. The formula (\ref{nad=0}) also shows that in the hyperplane $d=0$ the vector field
of the system (\ref{polykacz}) is parallel to the vector field of the system
(\ref{pierwotny}) and so their trajectories must coincide.
\end{proof}

Thus we have shown that the system (\ref{pierwotny})
is embedded in the completely integrable bi-Hamiltonian system (\ref{polykacz}). The trajectories
of (\ref{pierwotny}) stay on the intersection of invariant manifolds for (\ref{polykacz}) with
the hyperplane $d=0$. Also, since we can now solve the system (\ref{polykacz}) by quadratures
the time evolution of the coefficient $\lambda \det(B)$ in (\ref{nad=0}) can be calculated
which makes it possible to solve the system (\ref{pierwotny}) by quadratures too.

% Section on driven qLN systems - from Hans Lundmark

\setcounter{section}{7}
\setcounter{proposition}{0}

\section*{VII. New types of separation variables for driven qLN systems}

\setcounter{equation}{0}
\def\theequation{7.\arabic{equation}}

In this section we study an important class of two-dimensional qLN equations
called driven systems. We find for all such systems their separation variables
and prove their integrability by quadratures. The variables of separation
are of a completely new type: they consist of families of conics
which are non-confocal in contrast with the classical separability theory for potential systems.

We remind the reader that we call a two-dimensional Newton system {\em driven} 
if one of the two differential
equations depends on one variable only.
By renaming the variables if necessary, we can always arrange for
such a system to take the form
%\begin{align*}
%\ddot{q}_1 &= M_1(q_1,q_2)\\
%\ddot{q}_2 &= M_2(q_2)
%\end{align*}
\begin{equation}\label{driven7}
  \begin{array}{l}
    \ddot{q}_1 = M_1(q_1,q_2), \\
    \ddot{q}_2 = M_2(q_2).
  \end{array}
\end{equation}
The second equation can be solved on its own and
its solution $q_2(x)$ then determines the equation for $q_1$,
which explains the name ``driven''.
A driven system always has one integral of motion
$F=\dot{q}_2^2 /2 - \int M_2 \, dq_2$,
%\dot{q}^t \begin{bmatrix} 0 & 0 \\ 0 & 1 \end{bmatrix} \dot{q} -
%2\int M_2 \, dq_2, 
obtained by integrating the second equation once,
but in general there need not exist any others.

Here we shall consider driven systems that admit
a quasi-Lagrangian formulation 
$\ddot{q}=-\frac{1}{2}A^{-1}\nabla k(q)$.
Here, as usual,
$A(q)$ is a non-degenerate $2\times 2$ matrix satisfying the
cyclic conditions (\ref{cyclicn}),
i.e., a matrix of the form
\begin{equation}\label{explicit_A}
A=\left[
\begin{array}{cc}
a q_2^2+b q_2+\alpha &
-a q_1 q_2 - \frac{b}{2} q_1 - \frac{c}{2} q_2 + \frac{\beta}{2} \\
-a q_1 q_2 - \frac{b}{2} q_1 - \frac{c}{2} q_2 + \frac{\beta}{2} &
a q_1^2+c q_1+\gamma
\end{array}
\right].
\end{equation}
Such a system always has {\em two} functionally independent
integrals of motion
$E=\dot{q}^t A \dot{q} + k(q)$ and
$F=\dot{q}_2^2 /2 - \int M_2 \, dq_2$.

By examining the second component of the equation
$\ddot{q}=-\frac{1}{2}A^{-1}\nabla k(q)$,
we immediately see that a qLN system
is driven iff
%\begin{equation}
%\frac{-1}{2\det(A)}
%\left( -A_{12} \partial_1 k + A_{11} \partial_2 k \right)
%=M_2(q_2),
%\end{equation}
%or, equivalently,
\begin{equation}\label{driven_k}
A_{12} \partial_1 k - A_{11} \partial_2 k
=2 \det(A) M_2(q_2),
\end{equation}
for some function $M_2(q_2)$ depending on $q_2$ only.
We can produce driven qLN systems with any given
$M_2(q_2)$ and $A(q)$
by solving for $k(q)$ in this equation.
The case $A_{11}=0$ is degenerate and will be treated separately later
(see Remark \ref{case_A11=0}),
so we assume from now on that
$A_{11}\neq 0$.

We start by introducing  separation variables for (\ref{driven7})
as characteristic coordinates for (\ref{driven_k}).

\begin{definition}\label{def_uv}
Define curvilinear coordinates $(u,v)=(u(q),v(q))$ as follows.
Let $u$ be a parameter indexing
the family of characteristic curves of (\ref{driven_k}) given by
\begin{equation}\label{char_curve}
\dot{q}(x)=
\left[ \begin{array}{r}
A_{12}(q(x)) \\
-A_{11}(q(x))
\end{array} \right],
\end{equation}
and let $v=q_2$.
\end{definition}

In other words, the curves given by (\ref{char_curve}) are the
coordinate curves of constant $u$.
For a given matrix $A$ these curves can be explicitely calculated. 
In Theorem \ref{driven_coord_types} we will
describe these curves more explicitly.
Let us just note for the moment that
%the curves given by (\ref{char_curve})
they are not parallel to the curves of constant $v$,
because of the assumption $A_{11}\neq 0$.
Thus the above description really defines a coordinate system
(at least locally).
There is some freedom in the choice of $u$,
but this will not affect our results.
By abuse of notation we will write $f(q_1,q_2)$ and $f(u,v)$
for the same function $f$ expressed in different coordinate systems.

\begin{lemma}\label{driven_k_solution}
The general solution of (\ref{driven_k}) is
\begin{equation}
k(u,v) = f(u)+D(u,v)g(v),
\end{equation}
where $f$ is an arbitrary function,
$D=\det(A)$
and 
\begin{displaymath}
g(q_2)=\frac{-2}{A_{11}(q_2)} \int M_2(q_2) \, dq_2.
\end{displaymath}
\end{lemma}

\begin{proof}
%Let $q(x)$ be a characteristic curve given by (\ref{char_curve}).
%Along this curve we can consider (\ref{driven_k}) as an ODE
Along each characteristic curve $q(x)$ given by (\ref{char_curve})
we can consider (\ref{driven_k}) as an ODE
\begin{displaymath}
\frac{d}{dx}k(q(x)) = 2 D(q(x)) M_2(q_2(x)),
\end{displaymath}
with general solution
\begin{displaymath}
k(q(x)) = 
%\int 2 D(q(x)) M_2(q_2(x)) \,dx =
D(q(x))g(q_2(x)) + f,
\end{displaymath}
where $f$ is a constant of integration.
This can be verified by direct differentiation;
the cyclic conditions imply that
\begin{displaymath}
\frac{d}{dx} D(q(x)) = 
\partial_1 D \, \dot{q}_1 + \partial_2 D \, \dot{q}_2 =
\partial_1 D \, A_{12} + \partial_2 D \, (-A_{11}) =
-D \, \partial_2 A_{11},
\end{displaymath}
and thus
\begin{displaymath}
\begin{array}{rcl}
{\displaystyle\frac{d}{dx}} k(q) &=&
{\displaystyle\frac{d}{dx}} \left( D(q) g(q_2) + f \right) =
{\displaystyle\frac{dD}{dx}} \, g + D\, \partial_2 g\, \dot{q}_2 \\
&=& -D \, \partial_2 A_{11} \, g + D \, \partial_2 g \,(-A_{11}) \\
&=& -D \,\partial_2 (A_{11} g) = 2D M_2.
\end{array}
\end{displaymath}
The constant of integration $f$ can be different for different
characteristic curves,
so when we express the result in terms of $u$ and $v$,
$f$ will depend on $u$ (but not on $v$).
\end{proof}

\begin{lemma}
Equation (\ref{driven_k}) is equivalent, under the substitution
$k=K \det(A)$,
to the equation
\begin{equation}\label{fund_eqn_driven}
A_{12} \partial_{11}K -  A_{11} \partial_{12}K
-\frac{3}{2} \, \partial_2 A_{11} \, \partial _1 K = 0,
\end{equation}
which is the fundamental equation (\ref{fundeq}) associated with the matrices
$A$ and
\begin{displaymath}
B=\left[ \begin{array}{cc} 0 & 0 \\ 0 & 1/2 \end{array} \right].
\end{displaymath}

\end{lemma}

\begin{proof}
Equation (\ref{driven_k}) implies
\begin{displaymath}
\partial_1
\left( \frac{ -A_{12} \partial_1 k + A_{11} \partial_2 k}{\det(A)} \right)
=0.
\end{displaymath}
Conversely, this expression can be integrated to give (\ref{driven_k}),
where $M_2(q_2)$ is an arbitrary function of integration.
By substituting $k=K \det(A)$
and simplifying the resulting expression using the cyclic conditions
one obtains (\ref{fund_eqn_driven}).
Comparison with the general expression for the fundamental
equation in Theorem \ref{fundth} proves the second statement of the lemma.
\end{proof}

\begin{remark}\label{driven_remark}
The fundamental equation (\ref{fund_eqn_driven}) is hyperbolic.
Its characteristic coordinates are precisely
the coordinates $(u,v)$ of Definition \ref{def_uv}.
The general solution is $K(u,v)=f(u)/D(u,v)+g(v)$, as can be seen by
combining the above lemmas.
\end{remark}

Let us turn to the question of how
to integrate a driven qLN system. 
The solution $q_2(x)$ of the second equation can
be found by quadrature from $F=\dot{q}_2^2  /2 - \int M_2 \, dq_2$:
\begin{equation}
\int \frac{dq_2}{\sqrt{2F+2\int M_2 \, dq_2}} =
\pm \int dx.
\end{equation}
Inserting $q_2(x)$ and $\dot{q}_2(x)$ into
\begin{displaymath}
E=A_{11}(q_2)\,\dot{q}_1^2 + 2A_{12}(q_1,q_2)\,\dot{q}_1 \dot{q}_2 +
A_{22}(q_1)\,\dot{q}_2^2 + k(q_1,q_2)
\end{displaymath}
would give a first order ODE for $q_1(x)$,
but there is no obvious way to solve this equation
since the variables $q_1$ and $x$
do not separate.
We will now show how to proceed instead.

\begin{theorem}
%Every driven qLN system is separable in the characteristic coordinates
%of the fundamental equation (\ref{fund_eqn_driven}),
%and thus integrable by quadratures.
Every driven qLN system can be integrated by quadratures
using the characteristic coordinates $(u,v)$ of the fundamental equation
(\ref{fund_eqn_driven})
as separation variables.
\end{theorem}

\begin{proof}
We use the notation of Lemma \ref{driven_k_solution}.
Let the system be generated by $E=\dot{q}^t A \dot{q} + k(q)$
with $k(u,v) = f(u)+D(u,v)g(v)$. 
Since $v=q_2$,
we can express $F$ as $F=\frac{1}{2} \left( \dot{v}^2 + A_{11}(v)g(v) \right)$
and calculate $v(x)$ by quadrature, as above.
%Our aim now is to express 
%$E$ in terms of $u$ and $v$.
Now note that since the curves of constant $u$
by definition have tangent
$\dot{q}=(A_{12},-A_{11})^t$, we must have
$\nabla u = \rho(q) \, (A_{11},A_{12})^t$
for some function $\rho(q)$, whose exact form depends on the choice of $u$.
This gives
$\dot{u} = \partial_1 u \, \dot{q}_1 + \partial_2 u \, \dot{q}_2 =
\rho(q) \left( A_{11} \dot{q}_1 + A_{12} \dot{q}_2 \right)$,
and thus
\begin{eqnarray*}
\dot{u}^2 &=&
\rho^2 A_{11}
\left( A_{11} \dot{q}_1^2 + 2 A_{12} \dot{q}_1 \dot{q}_2 +
       \frac{A_{12}^2}{A_{11}} \dot{q}_2^2 \right) \\
&=& \rho^2 A_{11}
\left( E-A_{22}\dot{q}_2^2 - k(q) +
       \frac{A_{12}^2}{A_{11}} \dot{q}_2^2 \right) \\
&=& \rho^2 A_{11}
\left( E- \left( A_{22}-\frac{A_{12}^2}{A_{11}} \right) \dot{v}^2
  - f(u) - Dg(v) \right) \\
&=& \rho^2 A_{11}
\left( E-\frac{D}{A_{11}}\left(2F-A_{11}g(v) \right) - f(u) - Dg(v) \right) \\
&=& \rho^2 A_{11}
\left( E-\frac{2D}{A_{11}} F - f(u) \right).
\end{eqnarray*}
In order to complete the proof, we will show that
$\rho(u,v)=\phi(u) \, \left| A_{11}(v) \right|^{-3/2}$
and $D(u,v)/A_{11}(v) = \psi(u)$
for some functions $\phi$ and $\psi$,
since this implies that the variables $u$ and $x$ separate.
Explicitly, we can then find $u(x)$
from the quadrature
\begin{equation}\label{kwadratura}
\int \frac{du}{\phi(u) \sqrt{ E-2\psi(u)F-f(u) }} =
\pm \int \frac{dx}{A_{11}(v(x))},
\end{equation}
after which the inverse coordinate transformation gives us $q_1(x)$. Notice, that for a given matrix
$A$ the characteristic coordinates $(u,v)$ can be calculated explicitly so that the function $\rho$ and
thus $\phi$ and $\psi$ can be easily calculated and used in the quadrature (\ref{kwadratura}) above.
The theorem covers however all the cases at once without any need of calculating $\rho$ explicitly.

To see that $\rho(u,v)=\phi(u) \, \left| A_{11}(v) \right|^{-3/2}$,
note that $\partial_{12}u=\partial_{21}u$
implies that $\rho(q)$ satisfies the PDE
\begin{displaymath}
0=\partial_1 (\rho A_{12}) - \partial_2 (\rho A_{11}) =
A_{12} \partial_1 \rho - A_{11} \partial_2 \rho
- \frac{3}{2}\partial_2 A_{11} \, \rho,
\end{displaymath}
which has the same characteristic curves (\ref{char_curve})
as equation (\ref{driven_k}).
Along such a curve we determine $\rho$ by integrating
\begin{displaymath}
\frac{d}{dx}\rho(q(x)) = \frac{3}{2} \partial_2 A_{11}(q_2(x))\, \rho(q(x))
\end{displaymath}
which, taking into account $\dot{q}_2(x)=-A_{11}(q_2(x))$,  gives
\begin{displaymath}
\rho(q(x)) = \phi \, \left| A_{11}(q_2(x)) \right|^{-3/2}.
\end{displaymath}
The integration constant $\phi$ can be different on different
characteristic curves,
so changing to $(u,v)$ coordinates we obtain
\begin{displaymath}
\rho(u,v) = \phi(u) \, \left| A_{11}(v) \right|^{-3/2},
\end{displaymath}
as desired.

Finally, we calculate the total derivative of the function
$\psi(q)=D(q)/A_{11}(q_2)$
along a characteristic curve:
\begin{displaymath}
\frac{d}{dx}\psi(q(x))=
%\partial_1 \left( \frac{D}{A_11} \right) \dot{q}_1 +
%\partial_2 \left( \frac{D}{A_11} \right) \dot{q}_2 =
\partial_1 \!\left( A_{22}-\frac{A_{12}^2}{A_{11}} \right) A_{12} -
\partial_2 \!\left( A_{22}-\frac{A_{12}^2}{A_{11}} \right) A_{11}.
\end{displaymath}
Using the cyclic conditions,
we find that this expression is identically zero.
This implies that $\psi$ is constant along the
coordinate curves of constant $u$, i.e., $\psi=\psi(u)$.
%which is just what we wanted to show.
This completes the proof.
\end{proof}

\begin{remark}\label{case_A11=0}
The degenerate case $A_{11}=0$ can be treated as follows.
Since $a=b=\alpha=0$, the expression (\ref{explicit_A}) for $A$ reduces to
\begin{equation}
A=\left[
\begin{array}{cc}
0 &
-\frac{c}{2} q_2 + \frac{\beta}{2} \\
-\frac{c}{2} q_2 + \frac{\beta}{2} &
c q_1+\gamma
\end{array}
\right]
\end{equation}
and equation (\ref{driven_k}) reduces to
\begin{displaymath}
A_{12}\,\partial_1 k = 2 \left( -A_{12}^2 \right) M_2(q_2),
\end{displaymath}
with the general solution
\begin{displaymath}
k(q)=-2\,A_{12}(q_2)\, M_2(q_2)\,q_1 + k_2(q_2).
\end{displaymath}
We calculate $q_2(x)$ by quadrature as before.
Inserting $q_2(x)$ and $\dot{q}_2(x)$ into
$E= 2A_{12} \dot{q}_1 \dot{q}_2 + A_{22} \dot{q}_2^2 + k(q_1,q_2)$
yields in this case an equation of the form
$\dot{q}_1(x) + \xi(x) q_1(x) = \eta(x)$,
from which we can find $q_1(x)$ by quadrature.
\end{remark}

\begin{theorem}\label{driven_coord_types}
The separation coordinates for driven qLN systems,
i.e.,
the characteristic coordinates $(u,v)$ of the fundamental equation (\ref{fund_eqn_driven}),
are of one of the following types,
determined by the coefficients in the matrix $A$:
\begin{enumerate}
\item\label{fanlike}
Fan-like hyperbolic, if $a\neq 0$ and $b^2/4-a\alpha = 0$.
\item\label{axial_hyp}
Axial hyperbolic, if $a\neq 0$ and $b^2/4-a\alpha < 0$.
\item\label{2p_ell_hyp}
Two-point elliptic-hyperbolic, if $a\neq 0$ and $b^2/4-a\alpha > 0$.
\item\label{1p_par}
One-point parabolic, if $a=0$ and $b \neq 0$.
\item\label{par_par}
Parallel parabolic, if $a=0$ and $b = 0$.
\end{enumerate}
\end{theorem}

\begin{proof}
We will compute explicitly the curves given by (\ref{char_curve}),
which constitute the curves of constant $u$.
(The curves of constant $v$ are just horizontal lines, since $v=q_2$.)
Inserting the explicit expression (\ref{explicit_A})
for the matrix $A$ into (\ref{char_curve}),
we obtain
\begin{equation}\label{explicit_char_curve}
\left[ \begin{array}{l}
          \dot{q}_1 \\ \dot{q}_2
        \end{array} \right]
=
- \left[ \begin{array}{c}
       a q_1 q_2 + \frac{b}{2} q_1 + \frac{c}{2} q_2 - \frac{\beta}{2} \\
       a q_2^2+b q_2+\alpha 
        \end{array} \right].
\end{equation}
When solving these equations, we distinguish four different cases,
depending on the values of the parameters in $A$.

\medskip
{\it The case $a\neq 0$}.
\smallskip

By setting $r_1=a q_1+c/2$ and $r_2=a q_2+b/2$,
which is just rescaling of the axes and translation of the origin,
we transform (\ref{explicit_char_curve}) into
\begin{equation}
\left[ \begin{array}{l}
          \dot{r}_1 \\ \dot{r}_2
        \end{array} \right]
=
\left[ \begin{array}{c}
       -r_1 r_2 + C_1\\
       -r_2^2 +C_2
        \end{array} \right],
\quad{\rm where}\quad
\left\{ \begin{array}{l}
       C_1 = bc/4+a\beta/2 \\
       C_2 = b^2/4-a\alpha
\end{array}\right. .
\end{equation}

\medskip
{\it Subcase $C_2=0$ (type \ref{fanlike})}.
\smallskip

Either $r_2 = 0$,
or $r_2=(x+D_1)^{-1}$ and $r_1=C_1 (x+D_1)/2+ D_2(x+D_1)^{-1}$,
where $D_1$ and $D_2$ are constants of integration.
Eliminating $x$ and writing $u$ instead of $D_2$, we obtain
\begin{equation}
r_1 = \frac{C_1}{2 r_2} + u r_2,
\end{equation}
which represents a family of hyperbolas,
each with asymptotes $r_2=0$ and $r_2=r_1/u$.
The solution $r_2=0$ found above corresponds to the limiting cases
$u\to\pm\infty$ (see Figure 2).

\medskip
{\it Subcase $C_2\neq 0$ (types \ref{axial_hyp} and \ref{2p_ell_hyp})}.
\smallskip

The substitution $s_1=r_1 - C_1 r_2/C_2$, $s_2=r_2$ yields
\begin{equation}
\left[ \begin{array}{l}
          \dot{s}_1 \\ \dot{s}_2
        \end{array} \right]
=
\left[ \begin{array}{c}
       -s_1 s_2 \\
       -s_2^2 +C_2
        \end{array} \right],
\end{equation}
and thus
\begin{displaymath}
\frac{ds_1}{ds_2} = \frac{\dot{s}_1}{\dot{s}_2} = \frac{s_2}{s_2^2-C_2}s_1,
\end{displaymath}
resulting in
\begin{equation}
%s_1 = u \sqrt{ \left| s_2^2 - C_2 \right| }.
s_1^2 = u^2 \left| s_2^2 - C_2 \right| .
\end{equation}
If $C_2<0$ (type \ref{axial_hyp}),
this represents in the $s$-plane a family of hyperbolas centered around
the $s_1$-axis, with asymptotes $s_2=\pm s_1/u$ and vertices
$(\pm u\sqrt{-C_2},0)$ (Figure 3a).

If $C_2>0$ (type \ref{2p_ell_hyp}), we obtain in the region
$|s_2|>\sqrt{C_2}$
a family of hyperbolas
%centered around the $s_2$-axis,
with asymptotes $s_2=\pm s_1/u$
and vertices $(0,\pm\sqrt{C_2})$,
and in the region
$|s_2|<\sqrt{C_2}$
a family of ellipses with vertices
$(0,\pm\sqrt{C_2})$ and $(0,\pm u\sqrt{C_2})$ (Figure 4a).
The corresponding curves in the $r$-plane are obtained by a shear in the
$s_1$-direction with factor $C_1/C_2$ (Figures 3b and 4b).
They are still hyperbolas and ellipses, but not aligned parallel with the
$r$-axes.

\medskip
{\it The case $a=0$}.

\smallskip
{\it Subcase $b\neq 0$ (type \ref{1p_par})}.
\smallskip

Translating the origin by $r_1=q_1-(\alpha c/b^2+\beta/b)$
and $r_2=q_2+\alpha/b$,
we obtain
\begin{equation}
\left[ \begin{array}{l}
          \dot{r}_1 \\ \dot{r}_2
        \end{array} \right]
=
\left[ \begin{array}{c}
       -\frac{b}{2}r_1 -\frac{c}{2}r_2 \\
       -br_2
        \end{array} \right],
\end{equation}
which yields
\begin{equation}
\left( r_1 - \frac{c}{b}r_2 \right)^2 = u r_2.
\end{equation}
With $s_1=r_1-cr_2/b$,$s_2=r_2$,
we obtain in the $s$-plane a family of parabolas
$s_2=s_1^2/u$ (Figure 5a).
The corresponding curves in the $r$-plane are parabolas obtained
by a shear in the $s_1$-direction with factor $c/b$ (Figure 5b).

\medskip
{\it Subcase $b=0$ (type \ref{par_par})}.
\smallskip

Here we can assume that $\alpha\neq 0$, or else we get the degenerate
case $A_{11}=0$.
A simple calculation shows that
\begin{equation}
q_1 = -\frac{c}{4\alpha}q_2^2 + \frac{\beta}{2\alpha}q_2 + u,
\end{equation}
which is a family of translated parabolas seen on Figure 6 (or straight lines, if $c=0$).

\end{proof}

\enlargethispage{\baselineskip}

\begin{center}
\epsfig{file=fanlike.eps,height=4.5cm}
\end{center}

\vspace{5pt}

\begin{center}
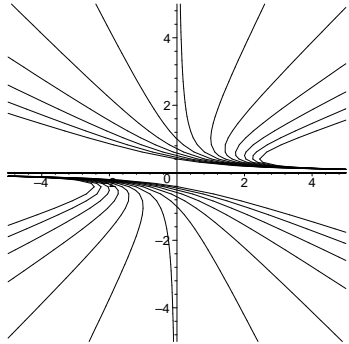
Figure 2. Fan-like hyperbolic.
\end{center}

\newpage

\begin{center}
\epsfig{file=axial_s.eps,height=8cm}
\end{center}

\vspace{20pt}

\begin{center}
Figure 3a. Axial-hyperbolic in the s-plane.
\end{center}

\begin{center}
\epsfig{file=axial_r.eps,height=8cm}
\end{center}

\vspace{20pt}

\begin{center}
Figure 3b. Axial-hyperbolic in the r-plane.
\end{center}

\newpage

\begin{center}
\epsfig{file=twopt_s.eps,height=8cm}
\end{center}

\vspace{20pt}

\begin{center}
Figure 4a. Two-point elliptic-hyperbolic in the s-plane.
\end{center}

\begin{center}
\epsfig{file=twopt_r.eps,height=8cm}
\end{center}

\vspace{20pt}

\begin{center}
Figure 4b. Two-point elliptic-hyperbolic in the r-plane.
\end{center}

\newpage

\begin{center}
\epsfig{file=onept_s.eps,height=8cm}
\end{center}

\vspace{20pt}

\begin{center}
Figure 5a. One-point parabolic in the s-plane.
\end{center}

\begin{center}
\epsfig{file=onept_r.eps,height=8cm}
\end{center}

\vspace{20pt}

\begin{center}
Figure 5b. One-point parabolic in the r-plane
\end{center}

\newpage

\begin{center}
\epsfig{file=parallel.eps,height=8cm}
\end{center}

\vspace{20pt}

\begin{center}
Figure 6. Parallel-parabolic.
\end{center}

% Section with examples of (2-dim) qLN systems

\setcounter{section}{8}
\setcounter{proposition}{0}

\section*{VIII. Examples and applications} % of qLN theory to given Newton equations}

\setcounter{equation}{0}
\def\theequation{8.\arabic{equation}}

The notion of a qLN force
$M(q)=-\frac{1}{2}A^{-1}(q) \nabla k(q)$
naturally generalizes the concept of a potential force
$M(q)=-\nabla k(q)$, which is a special case.
The qlN forces admit an integral of motion quadratic in velocities,
which in the potential case becomes the energy integral.
The function $k(q)$ may be called a ``quasi-potential'' of the force $M(q)$.

A given force is easy to test for the existence of a qLN formulation,
provided that one knows the general form of the matrix $A(q)$
solving the cyclic condition (\ref{cyclicn}).
In two dimensions $A(q)$, given by
\begin{displaymath}
A(r,w)=\left[
\begin{array}{cc}
a w^2+b w+\alpha &
-a r w - \frac{b}{2} r - \frac{c}{2} w + \frac{\beta}{2} \\
-a r w - \frac{b}{2} r - \frac{c}{2} w + \frac{\beta}{2} &
a r^2+c r+\gamma
\end{array}
\right],
\end{displaymath}
depends on 6 arbitrary parameters,
and a qLN formulation exists provided that the mixed derivatives of
$\nabla k(r,w) = -2A(r,w)M(r,w)$
are equal for some non-zero values of the parameters
$a$, $b$, $c$, $\alpha$, $\beta$, $\gamma$.
We thus have the following lemma:

\begin{lemma} \label{lemma:qln}
A given force $M(r,w)=(M_1(r,w), M_2(r,w))^t$ admits a qLN formulation
$M(r,w)=-\frac{1}{2}A^{-1}(r,w) \nabla k(r,w)$
if and only if
there is a non-trivial
solution $A$, with $\det(A) \ne 0$, of the equation
%\begin{equation}
%\frac{\partial}{\partial w}(AM)_1 =
%\frac{\partial}{\partial r}(AM)_2,
%\end{equation}
%or, explicitly,
\begin{eqnarray} 
0&=&\frac{\partial}{\partial w}
\left((a w^2+b w+\alpha) M_1 +
(-a r w - \frac{b}{2} r - \frac{c}{2} w + \frac{\beta}{2}) M_2 \right) -
\nonumber \\
&&\frac{\partial}{\partial r}
\left((-a r w - \frac{b}{2} r - \frac{c}{2} w + \frac{\beta}{2}) M_1 +
(a r^2+c r+\gamma) M_2 \right). \label{qln_condition}
\end{eqnarray}
\end{lemma}

\begin{lemma} {\bf (Criterion of integrability, n=2)} \label{lemma:2qln}
Equation (\ref{qln_condition}) has
a two-parameter family of solutions for $A(r,w)$
if and only if $\ddot q=M(q)$ admits two functionally
independent integrals of motion $E$ and $F$
quadratic in velocities.
\end{lemma}

\begin{proof}
If such $E=\dot q^t A \dot q + k$ and $F=\dot q^t B \dot q + l$
exist, then
$\lambda E +\mu F = \dot q^t (\lambda A +\mu B) \dot q + (\lambda k +\mu l)$
is an integral of motion for all $\lambda$ and $\mu$,
and thus $\lambda A +\mu B$ is a two-parameter solution of \ref{qln_condition}.

Conversely, if there is a two-parameter solution $D(\lambda,\mu)$ of \ref{qln_condition},
then there are linearly independent
integrals $E$ and $F$ with $A=D(1,0)$ and $B=D(0,1)$.
\end{proof}

These two lemmas make it simple to test a given two-dimensional
force for the existence of a qLN formulation,
and to show integrability if a two-parameter family of solutions
for $A$ exists.

\begin{example} {\bf (gH-H system)}.
The generalized H\'enon-Heiles (gH-H) system \cite{biham_HH}
is defined by the potential
\begin{displaymath}
V(q_1,q_2) = c_1 q_1 q_2^2-\frac{c_2 q_1^3}{3}+\frac{c_0}{2 q_2^2},
\quad c_1,c_2\neq 0.
\end{displaymath}
It is known to be integrable in three cases:
the Korteweg-de Vries (KdV) case $6 c_1+c_2=0$,
the Sawada-Kotera (S-K) case $c_1+c_2=0$
and the Kaup-Kupershmidt (K-K) case $16 c_1+c_2=0$.
In the KdV case, and also in the S-K case if $c_0=0$,
the second integral of motion is quadratic in velocities,
in the other cases it is quartic.
The system appears naturally when integrating the equation
$0=(\frac14\partial^3 + 2c_0 u\partial + c_0 u_x)
(\frac14 u_{xx}-\frac14 c_2 u^2)$,
%third-order Hamiltonian operator
%$\frac{1}{4}\partial^3 + 2c_0 u\partial + c_0 u_x$
which for the above cases corresponds to
the stationary flow of the fifth order KdV, S-K and K-K soliton equations.
This observation explained \cite{fordy} the remarkable connection
of the integrable cases of the gH-H system with soliton hierarchies.

We shall apply our criterion for existence of a qLN formulation to two
Newton representations of the gH-H system; 
the original system in $q$-variables
\begin{eqnarray} \label{gHHq}
&& \ddot q_1 = -\frac{\partial V}{\partial q_1} =
            -c_1 q_2^2 + c_2 q_1^2, \nonumber \\
&& \ddot q_2 = -\frac{\partial V}{\partial q_2} =
            -2c_1 q_1 q_2 + \frac{c_0}{q_2^3},
\end{eqnarray}
and another system in $r$-variables
\begin{eqnarray} \label{gHHr}
&& \ddot r_1 = r_2 + (c_1+c_2)r_1^2, \nonumber \\
&& \ddot r_2 = c_3-10 c_1 r_1 r_2 - 10 c_1\left(c_1+\frac{c_2}{3}\right) r_1^3,
\end{eqnarray}
%These systems are equivalent under the map
which is equivalent \cite{gHH} to the $q$-system under the map
$r_1=q_1$, $r_2=-c_1 (q_1^2+q_2^2)$,
$c=-4 c_1 \left[\frac{1}{2}(\dot q_1^2+\dot q_2^2)+V(q_1,q_2) \right]$.
The $r$-system does not have any natural Lagrangian or
Hamiltonian formulation
and its integrability has previously been studied only through its
equivalence with the original gH-H system.
We will show here a more direct approach based on the qLN theory.

Beginning with the $q$-system,
we insert the right-hand side $M$ from (\ref{gHHq})
into (\ref{qln_condition}),
identifying $(q_1,q_2)$ with $(r,w)$ as usual.
Since the powers $r^i w^j$ are linearly independent,
the coefficients at different powers must be individually zero.
This gives, after some simplification, that
$a=b=0$, $\alpha=\gamma$ arbitrary, $(6c_1+c_2)c=0$, $(c_1+c_2)\beta=0$
and $c_0 \beta=0$.
This means that we always have a solution
$A=tI$ (of course corresponding to the energy integral,
since the system has a potential).
Moreover, in two cases there exists a two-parameter solution;
when $6c_1+c_2=0$,
\begin{displaymath}
A=t \left[ \begin{array}{cc} 1 & 0 \\ 0 & 1 \end{array} \right]+
   s \left[ \begin{array}{cc} 0 & -q_2 \\ -q_2 & 2q_1 \end{array} \right]
\end{displaymath}
and when $c_1+c_2=c_0=0$,
\begin{displaymath}
A=t \left[ \begin{array}{cc} 1 & 0 \\ 0 & 1 \end{array} \right]+
   s \left[ \begin{array}{cc} 0 & 1 \\ 1 & 0 \end{array} \right].
\end{displaymath}
So in this way we have recovered the KdV and S-K ($c_0=0$) cases,
while the K-K and S-K ($c_0\neq 0$) cases,
with a quartic extra integral, fall outside of the qLN theory.

Performing the same procedure for the $r$-system (\ref{gHHr}),
we find that
$a=0$, $4c_1 \gamma+b=0$, $(31c_1+c_2)c=(3c_1^2+c_1 c_2)c=0$,
$2\alpha-3c_3 c=0$ and $(6c_1+c_2)\beta=0$.
Since we have excluded the trivial case $c_1=0$, it follows that
$c=\alpha=0$, so that the solution is
\begin{displaymath}
A=t \left[ \begin{array}{cc} -2 r_2 & r_1 \\  r_1 & 1/2c_1 \end{array} \right]
\end{displaymath}
except for the KdV case $6c_1+c_2=0$ which admits a two-parameter solution
\begin{displaymath}
A=t \left[ \begin{array}{cc} -2 r_2 & r_1 \\  r_1 & 1/2c_1 \end{array} \right]+
   s \left[ \begin{array}{cc} 0 & 1 \\ 1 & 0 \end{array} \right].
\end{displaymath}
This agrees with the known fact \cite{gHH} that for the $r$-system
the second integral is quartic in velocities in the S-K and K-K cases
and thus cannot be found by this method.

Suppose, however, that we had found the second integral in these cases
by some other method.
Then we would still have use for the qLN theory in 
proving the system's integrability,
since the $r$-paramerization admits
the nonstandard Hamiltonian formulation (\ref{hamzanurz})
with $\lambda=1$ and
$c_3$ playing the role of the fifth variable $d$. 
(This coincides with the Hamiltonian formulation that was found in \cite{gHH}
by transferring the standard Hamiltonian formulation from the
$q$-parametrization,
except for naming the momenta in reverse order;
here $p_i=\dot r_i$, while in that paper $s_1=\dot r_2$, $s_2=\dot r_1$.)

For example, in the S-K case ($c_1=-c_2=1/2$)
we have
\begin{displaymath} % \renewcommand{\arraystretch}{1.3}
 \left[ \begin{array}{c} \dot r_1 \\ \dot r_2 \\ \dot p_1  \\ \dot p_2 \\ \dot c_3 \end{array} \right]
 = \left[ \begin{array}{cccc|c}
          0 & 0 & -1/2 & -r_1/2 & p_1 \\
          * & 0 & -r_1/2 & r_2 & p_2 \\
          * & * & 0 & -p_1/2 & r_2 \\
          * & * & * & 0 & c_3 - 5 r_1 r_2 - \frac53 r_1^3 \\
 \hline
          * & * & * & * & 0 \end{array} \right] \nabla_{\!\cal M}c_3,
\end{displaymath}
with the commuting integrals of motion
$E=-2r_2 p_1^2 +2r_1 p_1 p_2 +p_2^2 +4r_1 r_2^2 +\frac23 r_1^3 (r_1^2+5r_2) + c_3 (-r_1^2-2r_2)$,
which is a Casimir, %of $\Pi$,
the Hamiltonian $c_3$,
and
$F=\frac32 p_1^4 - 6 r_1 r_2 p_1^2 + (3 r_1^2 - r_2) p_1 p_2 + r_1 p_2^2 +
\frac72 r_1^2 r_2^2 + \frac{10}{3} r_1^4 r_2 + \frac56 r_1^6 + \frac13 r_2^3
+ (-2 r_1 r_2 - r_1^3 + \frac32 p_1^2) c_3$, which is quartic in momenta.
\end{example}

\begin{example} {\bf (Quadratic force)}
As an example of direct application of the criterions in Lemma
\ref{lemma:qln} and Lemma \ref{lemma:2qln},
let us determine when the simple homogeneous quadratic force system
\begin{eqnarray}
&& \ddot r = \theta_1 r^2 + w^2, \nonumber \\
&& \ddot w = r^2 + \theta_2 w^2   \nonumber
\end{eqnarray}
admits two quadratic integrals of motion,
so that it is integrable in the sense explained previously.
Plugging into equation (\ref{qln_condition}) and equating the coefficients 
of independent powers $r^i w^j$ to zero,
we find $a=0$, $5\theta_1 b-7c=0$, $7b-5\theta_2 c=0$, $-\theta_2 b + \theta_1 c=0$,
$\theta_1 \beta + 2 \gamma =0$, $2\alpha +\theta_2 \beta =0$.
The two last equations always give rise to a one-parameter solution for
$\alpha$, $\beta$, $\gamma$,
while the equations involving $b$ and $c$ admit a
(real-valued) one-parameter solution
if and only if $\theta_1=\theta_2=7/5$,
which thus is the only case admitting a
two-parameter solution for the system as a whole.
The solution is in this case
\begin{displaymath}
A=t \left[ \begin{array}{cc} w & -(r+w)/2 \\  -(r+w)/2 & r \end{array} \right]+
   s \left[ \begin{array}{rr} -7 & 5 \\ 5 & -7 \end{array} \right].
\end{displaymath}
Integrating $\nabla k=-2AM$ we find the
two independent integrals of motions
$E=w\dot r^2 - (r+w) \dot r \dot w + r\dot w^2 +
(5r^4-12r^3w+14r^2w^2-12rw^3+5w^5)/28$
and
$F=-7\dot r^2 +10 \dot r \dot w + -7 \dot w^2 + 4(r^3+w^3)/5$.
\end{example}

The next example illustrates how an application of our
recursion formula (\ref{rekursjam})
reproduces a known \cite{ramani} family of potentials
separable in parabolic coordinates.
%similar to the Jacobi family of elliptic separable potentials
%(cf.~example \ref{jacobifam}).

\begin{example} {\bf (Parabolic separable potentials)}
The family of integrable potentials
\begin{displaymath} \arraycolsep 0pt 
V_n(r,w) = \sum_{k=0}^{\left[ n/2 \right]} 2^{n-2k} 
\left( \begin{array}{c} n-k \\ k \end{array} \right)
%\binom{n-k}{k} 
r^{2k} w^{n-2k},
\end{displaymath}
with second integral $I_n = -w \dot r^2 + r \dot r \dot w + r^2 V_{n-1}$,
was found in \cite{ramani}. 
These potentials are all separable \cite{ankiewicz} in parabolic coordinates
$r=\xi\eta$, $w=(\xi^2-\eta^2)/2$
%$\lambda_{1,2}=-w \pm \sqrt{r^2+w^2}$
by the Bertrand-Darboux theorem
and include
$V_3=4r^2 w+8w^3$, which is the KdV case of the H\'enon-Heiles potential,
as well as $V_4=r^4+12 r^2 w^2+16 w^4$, the so-called ``1:12:16-potential.''
We identify the matrices
\begin{displaymath}
A=\left[ \begin{array}{cc} -w & r/2 \\ r/2 & 0 \end{array} \right], \quad
B=\left[ \begin{array}{cc} 1/2 & 0 \\ 0 & 1/2 \end{array} \right]
\end{displaymath}
in the integrals of motion,
and take
$k_1 = r^2 V_0 = r^2$,
$l_1 = V_1 = 2w$.
It is then easy to show by induction that our recursion formula (\ref{rekursjam}),
which in this case reads
$k_m = -r^2 l_{m-1}$, $l_m = -2w l_{m-1} - k_{m-1}$,
reproduces this family up to a sign:
$(-1)^{m-1} k_m = r^2 V_{m-1}$, $(-1)^{m-1} l_m = V_m$, for all $m\ge 1$.
\end{example}

In order to give an impression of the wealth of different types of
non-potential Newton forces belonging to our theory,
we will now examine solutions of the fundamental equation (\ref{fundeq})
for some specified pairs of matrices $A$ and $B$.
Any such solution corresponds to an integrable qLN system,
and once one solution has been found,
a whole family of solutions can be constructed using the
recursion theorem \ref{recursionthm}.

\begin{example} {\bf (One-dimensional complex motion)}
If we take $A$ and $B$ as
\begin{displaymath}
A=\left[ \begin{array}{cc} 1 & 0 \\ 0 & -1 \end{array} \right],~~~~
B=\left[ \begin{array}{cc} 0 & 1 \\ 1 & 0 \end{array} \right],
\end{displaymath}
then the fundamental equation reduces to the Laplace equation
$K_{rr}+K_{ww}=0$.
Given a solution $K(r,w)$, i.e.\ a harmonic function,
we have $k=K \det(A)=-K$, so $k$ is also harmonic.
We find the corresponding $l$ from the relation
\begin{displaymath}
\nabla l = BA^{-1}\nabla k =
\left[ \begin{array}{cc} 0 & -1 \\ 1 & 0 \end{array} \right] \nabla k,
\end{displaymath}
which is nothing but the Cauchy-Riemann equations for $k$ and $l$,
so $l$ is the harmonic conjugate of $k$.
The corresponding qLN system $\ddot r=-k_r/2$, $\ddot w=k_w/2$
can be integrated by introducing
the complex variable $z=r+iw$ and the complex integral of motion
${\cal E} = E+iF = (\dot r^2 - \dot w^2 + k) + i(2\dot r \dot w + l) =
(\dot r+i \dot w)^2 + (k+il) = \dot z^2 + f(z)$,
where $f(z) = k(z) + i l(z)$ is analytic.
We can now determine $z$, and thus $r$ and $w$,
from $\dot z = \pm \sqrt{{\cal E}-f(z)}$
by one complex quadrature.

Repeated application of the recursion formula (\ref{rekursjam})
yields in this case the standard cycle of conjugate harmonic pairs
$(k,l)\to (l,-k)\to (-k,-l)\to (-l,k)\to (k,l)$.
\end{example}

\begin{example} {\bf (Fundamental equation separable in polar coordinates)}
Let
\begin{displaymath}
A=\left[ \begin{array}{cc} -2w & r \\ r & 0 \end{array} \right],~~~
B=\left[ \begin{array}{cc} 0 & w \\ w & -2r \end{array} \right].
\end{displaymath}
Then the fundamental equation becomes
$0=2(r^2 K_{rr} + 2rw K_{rw} + w^2 K_{ww}) + 9(rK_r+wK_w) + 6K$,
which in polar coordinates
($r=R\cos\phi$, $w=R\sin\phi$)
transforms into
$0=2R^2 K_{RR} + 9 R K_R + 6K$.
The general solution of this equation is
$K(R,\phi)=f_0(\phi)R^{-2} + g_0(\phi)R^{-3/2}$,
for some arbitrary functions $f_0$ and $g_0$.
Changing back to $r$ and $w$,
we find that the general solution of the fundamental equation in this case is
\begin{eqnarray}
K(r,w)&=&f_0\left( \arctan\frac{w}{r} \right) \frac{r^{-2}}{1+(w/r)^2} +
         g_0\left( \arctan\frac{w}{r} \right) \frac{r^{-3}}{\left(1+(w/r)^2\right)^{3/2}} \nonumber \\
&=& f\left( \frac{w}{r} \right) r^{-2} + g\left( \frac{w}{r} \right) r^{-3}, \nonumber
\end{eqnarray}
where $f$ and $g$ are arbitrary functions.
\end{example}

\begin{example}
Let
\begin{displaymath}
A=\left[ \begin{array}{cc} -w^2+a_1 & rw \\ rw & -r^2+a_2 \end{array} \right],~~~
B=\left[ \begin{array}{cc} b_1 & 0 \\ 0 & b_2 \end{array} \right]
\end{displaymath}
for some (non-zero) constants $a_1$, $a_2$, $b_1$, $b_2$.
Then the fundamental equation becomes
$0=-b_2 rw K_{rr} + (b_1 r^2-b_2 w^2+b_2 a_1-b_1 a_2) K_{rw} + c_1 rw K_{ww}
   -3b_2 w K_r + 3b_1 r K_w$.
Finding the general solution seems difficult in this case,
but looking for a particular solution of the form
$K(r,w)=f(r)+g(w)$, we find
\begin{displaymath}
K(r,w) = c_0 (b_1 r^2 +b_2 w^2) + \frac{c_1}{r^2} + \frac{c_2}{w^2} + c_3,
\end{displaymath}
where the $c_i$ are arbitrary constants.
To simplify formulas we let $c_0=1$, $c_1=c_2=c_3=0$, which gives 
$K_0(r,w) = b_1 r^2 +b_2 w^2$.
We take $l_1=K_0 \det(B)=b_1 b_2 K$ and determine $k_1$ from
$\nabla k_1=AB^{-1}\nabla l_1$, which gives
$k_1(r,w) =  b_1 b_2 (a_1 r^2 + a_2 w^2)$.
%        +\frac{c_1 b_2}{r^2} (a_1-w^2)
%        +\frac{c_2 b_1}{w^2} (a_2-r^2).
The corresponding qLN system
$\ddot q=-\frac12 A^{-1}\nabla k_1=-\frac12 B^{-1}\nabla l_1$
reads
%\begin{eqnarray}
%\ddot r &=& b_2 \left(- c_0 b_1 r + \frac{c_1}{r^3} \right), \nonumber \\
%\ddot w &=& b_1 \left(- c_0 b_2 w + \frac{c_2}{w^3} \right).
%\end{eqnarray}
\begin{eqnarray}
\ddot r &=& -b_1 b_2 r, \nonumber \\
\ddot w &=& -b_1 b_2 w. \nonumber
\end{eqnarray}
This system is separated in $r$ and $w$, and therefore trivially integrable.
However, we can obtain a more interesting integrable system by using
the recursion formula (\ref{rekursja}),
which here gives
\begin{eqnarray}
k_2 &=& \frac{a_1 a_2 - a_1 r^2 - a_2 w^2}{b_1 b_2}\, l_1 \nonumber \\
%    &=& (a_1 a_2 - a_1 r^2 - a_2 w^2)
%        \left( c_0 (b_1 r^2 +b_2 w^2) + \frac{c_1}{r^2} + \frac{c_2}{w^2} + c_3 \right), \nonumber \\
    &=& (a_1 a_2 - a_1 r^2 - a_2 w^2) (b_1 r^2 +b_2 w^2), \nonumber \\
l_2 &=& \frac{a_1 b_2 + a_2 b_1 - b_1 r^2 - b_2 w^2}{b_1 b_2}\, l_1 - k_1 \nonumber \\
%    &=&  c_0 (a_2 b_1^2 r^2 + a_1 b_2^2 w^2 - (b_1 r^2 + b_2 w^2)^2)+
%         c_1 b_1 \frac{a_2 - r^2}{r^2} + \\
%&&       c_2 b_2 \frac{a_1 - w^2}{w^2} +
%         c_3 (a_1 b_2 + a_2 b_1 - b_1 r^2 - b_2 w^2).
    &=&  a_2 b_1^2 r^2 + a_1 b_2^2 w^2 - (b_1 r^2 + b_2 w^2)^2. \nonumber
\end{eqnarray}
The corresponding integrable qLN system
$\ddot q=-\frac12 A^{-1}\nabla k_2=-\frac12 B^{-1}\nabla l_2$
is
\begin{eqnarray}
%\ddot r &=& c_0 (2b_1 r^3 + 2b_2 rw^2 - a_2 b_1 r) +
%            \frac{c_1 a_2}{r^3} + c_3 r, \nonumber \\
%\ddot w &=& c_0 (2b_2 w^3 + 2b_1 wr^2 - a_1 b_2 w) +
%            \frac{c_2 a_1}{w^3} + c_3 w.
\ddot r &=& 2b_1 r^3 + 2b_2 rw^2 - a_2 b_1 r, \nonumber \\
\ddot w &=& 2b_2 w^3 + 2b_1 wr^2 - a_1 b_2 w. \nonumber
\end{eqnarray}
We derive from this also another solution
$K_1=l_2 / \det(B) =l_2 / b_1 b_2$
of the fundamental equation,
not of the form $K_1=f(r)+g(w)$.
Having found $K_0$ and $K_1$, we can employ the two-step recursion
(\ref{rekursjaK}),
which in this case reads
\begin{displaymath}
K_{m+1} = \frac{a_1 b_2 + a_2 b_1 - b_1 r^2 - b_2 w^2}{b_1 b_2}\, K_m
        - \frac{a_1 a_2 - a_1 r^2 - a_2 w^2}{b_1 b_2}\, K_{m-1},
\end{displaymath}
to find a whole sequence
$\left\{K_m \right\}_{-\infty}^{\infty}$ of solutions.
The first few are
\begin{eqnarray}
K_{-1} & = & \frac{b_1 b_2 (a_1 r^2 + a_2 w^2)}{a_1 a_2 - a_1 r^2 - a_2 w^2}, \nonumber \\
K_0 & = & b_1 r^2 +b_2 w^2, \nonumber \\
K_1 & = & \frac{a_2 b_1^2 r^2 + a_1 b_2^2 w^2 - (b_1 r^2 + b_2 w^2)^2}{b_1 b_2}, \nonumber \\
K_2 & = &  \big( (b_1 r^2 +b_2 w^2)^3
	+(a_2 b_1^2 r^2 + a_1 b_2^2 w^2)(a_1 b_2 + a_2 b_1 - b_1 r^2 - b_2 w^2)
\nonumber \\
&&	-(a_2 b_1^2 r^2 + a_1 b_2^2 w^2 + a_1 a_2 b_1 b_2)(b_1 r^2 +b_2 w^2)
	\big)/
	b_1^2 b_2^2. \nonumber
\end{eqnarray}
\end{example}

Finally, let us conclude with an example of a three-dimensional qLN system.
A detailed treatment of higher-dimensional
qLN systems is presented in a separate article \cite{hans}.
\begin{example}
The Newton system
\begin{eqnarray}
\ddot r_1 &=& -10 r_1^2 + 4 r_2, \nonumber \\
\ddot r_2 &=& -16 r_1 r_2 + 10 r_1^3 + 4 r_3, \nonumber \\
\ddot r_3 &=& -20 r_1 r_3 -8 r_2^2 + 30 r_1^2 r_2 - 15 r_1^4 + d \nonumber
\end{eqnarray}
was found in \cite{newton} as a parametrization of the 7th order
stationary KdV flow. 
It has three integrals of motion
\begin{eqnarray}
E_1 &=& \dot r_1 \dot r_3 + \frac{\dot r_2^2}{2} +
       10 r_1^2 r_3 - 4 r_2 r_3 + 8 r_1 r_2^2 - 10 r_1^3 r_2 + 3 r_1^5 - d r_1, \nonumber \\
E_2 &=& r_3 \dot r_1^2 - r_1 \dot r_2^2 + r_2 \dot r_1 \dot r_2 -
        \dot r_2 \dot r_3 - r_1 \dot r_1 \dot r_3 + \nonumber \\
    & & 4 r_1^2 r_2^2 + 5 r_1^4 r_2 - \frac52 r_1^6 - 4 r_2^3 +
        2 r_3^2 - 12 r_1 r_2 r_3 + \frac{d r_1^2}{2} + d r_2,\nonumber \\
E_3 &=& \frac18 (r_2^2 \dot r_1^2 + r_1^2 \dot r_2^2 +\dot r_3^2
        -(2r_1 r_2+4r_3)\dot r_1 \dot r_2 + 2r_1 \dot r_2 \dot r_3
        + 2r_2 \dot r_1 \dot r_3) +  \nonumber \\
&&      r_1 r_2^3 - 3 r_1^3 r_2^2 + \frac54 r_1^5 r_2 + 2r_1 r_3^2
        +\frac54 r_1^4 r_3 - r_1^2 r_2 r_3 + r_2^2 r_3
        -\frac{d}{4}(r_1 r_2 - r_3), \nonumber
\end{eqnarray}
which all are quadratic in velocities.
This means that the system is generated by any of them through
the quasi-Lagrangian equations.
From the velocity-dependent parts we find 
\begin{displaymath}
\left[ \begin{array}{ccc}
0 & 0 & 1 \\
0 & 1 & 0 \\
1 & 0 & 0 
\end{array} \right],
\,
\left[ \begin{array}{ccc}
2r_3 & r_2 & -r_1 \\
r_2 & -2r_1 & -1 \\
-r_1 & -1 & 0 
\end{array} \right],
\,
\left[ \begin{array}{ccc}
r_2^2 & -r_1r_2-2r_3 & r_2 \\
-r_1r_2-2r_3 & r_1^2 & r_1 \\
r_2 & r_1 & 1 
\end{array} \right]
\end{displaymath}
as examples of $3\times 3$-matrices satisfying the cyclic conditions (\ref{cyclicn}).
\end{example}

\setcounter{section}{9}
\setcounter{proposition}{0}

\section*{IX. Conclusions}

In this article we have developed a new theory --- the theory of quasi-Lagrangian
Newton equations. It was originally inspired  by interesting properties of the 
second stationary flow of the Harry Dym hierarchy, which led us to a broad theory
which encompasses the classical separability theory but 
goes far beyond the classical results --- the classical Bertrand-Darboux theory of
separability for two-dimensional potential forces depends on one essential free parameter
while our theory depends on five parameters. 

The main part of this work has been focused on two-dimensional qLN systems which admit
two integrals of motion $E$ and $F$ quadratic in velocities. These systems have only
a non-standard Hamiltonian formulation and are completely integrable by embedding
into five-dimensional Liouville integrable systems. All such qLN systems are characterized by 
a single PDE called here the fundamental equation. We have shown that there is a one-to-one
correspondence between fundamental equations and linear pencils $\lambda A + \mu B$ of matrices $A$
and $B$. These linear pencils have been classified in Sec. 5. 
In Sec. 7 the class of driven systems has been studied in detail and new types of 
separation variables (non-confocal conics) have been found. We have also shown that 
any given force can be effectively tested for the existence of qLN formulation, which can further
be used for unveiling its complete integrability and for solving the corresponding Newton
equation. We have illustrated this by several examples including the generalized H\'{e}non-Heiles
system (Sec. 8).

There are several natural directions of development of the theory of qLN systems. $n$-dimensional
versions of our main theorems on fundamental equation and on complete integrability
have already been formulated and proved in \cite{hans}.

The great wealth of different types of integrable Newton equations contained in the fundamental
equation remains to be studied. Here we have only discussed two special cases: separable systems
and driven systems. However, one of the most challenging questions yet to be answered
is the existence of separation variables for the fundamental equation in its most general form.
It can lead to new and interesting connections with the classical theory of separability
of the Hamilton-Jacobi equation and of linear PDE's.

%BIBLIOGRAPHY


\begin{thebibliography}{99}
\bibitem{two_newton}
 \art{S. Rauch-Wojciechowski, K. Marciniak, M. Blaszak}{Two Newton decompositions of stationary
  flows of KdV and Harry Dym hierarchies}{Physica A}{233}{1996}{307--330}
\bibitem{hans}
 {\sc H. Lundmark}, in: {\em Integrable Nonconservative Newton Systems with Quadratic Integrals of Motion},
Link\"{o}ping Studies in Science and Technology. Theses No. 756, Link\"{o}ping 1999.
\bibitem{whittaker}
 \book{E. T. Whittaker}{A treatise on the analytic dynamics of particles
              and rigid bodies}{Cambridge University Press, Cambridge, 1959, sec. 152}
\bibitem{garnier}
 \art{R. Garnier}{}{Circolo Mat. Palermo}{43}{1919}{155}
\bibitem{jacobi}
 \art{S. Wojciechowski}{Integrable one-particle potentials related to the Neumann system 
      and the Jacobi problem of geodesic motion on an ellipsoid}{Phys. Lett. A}
      {107}{1985}{106--111}
\bibitem{ankiewicz}
  \art{A. Ankiewicz, C. Pask}{The complete Whittaker theorem for two-di\-men\-sio\-nal
       integrable systems and its application}{J. Phys. A: Math. Gen.}{16}{1983}{4203--4208}
 \bibitem{Paper5}
  \art{K. Marciniak, S. Rauch-Wojciechowski}{Two families of non-\-stan\-dard Poisson structures
       for Newton equations}{J. Math. Phys.}{39}{1998}{5292--5306}
 \bibitem{seppoiss}
  \art{S. Rauch-Wojciechowski}{A bi-Hamiltonian formulation for separable potentials
       and its application to the Kepler problem and the Euler problem of two centers 
       of gravitation}{Phys. Lett. A}{160}{1991}{149--154}
 \bibitem{arnold}
  \book{V. I. Arnold}{Mathematical Methods of Classical Mechanics}
       {New York Berlin Springer-Verlag cop. 1989}
 \bibitem{biham_HH}
   \art{M. Antonowicz, S. Rauch-Wojciechowski}{Bi-Hamiltonian formulation of the 
          H\'enon-Heiles system and its multidimensional extensions}{Phys. Lett. A}{163}{1992}
          {167--172}
 \bibitem{fordy}
   \art{A. P. Fordy}{The {H}\'enon-{H}eiles system revisited}{Physica D}{52}{1991}{204--210}
 \bibitem{gHH}
    \art{M. B{\l}aszak, S. Rauch-Wojciechowski}{A generalized {H}\'enon-{H}eiles 
         system and related integrable Newton equations}{J. Math. Phys.}{35}{1994}{1693--1709}
 \bibitem{ramani}
   \art{A. Ramani, B. Dorizzi, B. Grammaticos}{Painlev\'e conjecture revisited}{Phys. Rev. Lett.}{49}
        {1982}{1539--1541}
 \bibitem{newton}
    \art{S. Rauch-Wojciechowski}{Newton representation for stationary flows of the KdV
             hierarchy}{Phys. Lett. A}{170}{1992}{91--94}
\end{thebibliography}
\end{document}